\documentclass[11pt,a4paper]{article}
\usepackage{jheppub}
\usepackage[utf8]{inputenc}
\usepackage{graphicx}
\usepackage{verbatim}
\usepackage{epsfig}
\usepackage{subfigure}
\usepackage{color}
\usepackage{multirow}
\usepackage{comment}
\usepackage{slashed}
\usepackage{amsmath}
\usepackage{ulem}
\usepackage{framed}

\usepackage{amssymb,amsmath,amsfonts,mathrsfs}
\usepackage[dvipsnames]{xcolor}
\usepackage{graphicx}
\usepackage{longtable}
\usepackage{verbatim}
\usepackage{color}
\usepackage[mathscr]{euscript}
\usepackage{framed}
\usepackage{mdframed} 

\usepackage{cancel}
\usepackage{color}
\usepackage{hyperref}
\hypersetup{colorlinks, citecolor=bluscuro, linkcolor=black, urlcolor=bluscuro}
\definecolor{rossos}{cmyk}{0,1,1,0.55}
\definecolor{bluscuro}{rgb}{0.15, 0.2, .85}
\definecolor{bluchiaro}{cmyk}{1,.3,0.,0.1}

\numberwithin{equation}{section}

\DeclareMathAlphabet\mathbfcal{OMS}{cmsy}{b}{n}

\usepackage{tikz}

\usepackage{tcolorbox}

\newcommand{\pt}[1]{\left(#1\right)}

\newcommand{\pmat}{\begin{pmatrix}}
\newcommand{\fpmat}{\end{pmatrix}}
\newcommand{\eq}{\begin{equation}}
\newcommand{\feq}{\end{equation}}
\newcommand{\cas}{\begin{cases}}
\newcommand{\fcas}{\end{cases}}

\newcommand{\eqarray}{\begin{eqnarray}}
\newcommand{\feqarray}{\end{eqnarray}}

\newcommand{\be}{\beta}

\newcommand{\ve}{\varepsilon}

\newcommand{\half}{\frac{1}{2}}

\def\ve{\epsilon}

\newcommand{\ft}[2]{{\textstyle\frac{#1}{#2}}}
\newcommand{\nn}{\nonumber}

\def\be{\begin{equation}}
\def\ee{\end{equation}}
\def\bea{\begin{eqnarray}}
\def\eea{\end{eqnarray}}

\usepackage{tensor}

\title{The multipolar structure of fuzzballs}
\author[a]{Massimo Bianchi,}
\author[b]{Dario Consoli,}
\author[a]{Alfredo Grillo,}
\author[a]{Jos\`e Francisco Morales,}
\author[c]{Paolo Pani,}
\author[c]{Guilherme Raposo.}

\affiliation[a]{Dipartimento di Fisica,  Università di Roma ``Tor Vergata"  \& Sezione INFN Roma2, Via della ricerca 
scientifica 1, 00133, Roma, Italy}
\affiliation[b]{Mathematical Physics Group, University of Vienna, Boltzmanngasse 5 1090 Vienna, Austria}
\affiliation[c]{Dipartimento di Fisica, ``Sapienza" Universit\`a di Roma \& Sezione INFN Roma1, Piazzale Aldo Moro 
5, 00185, Roma, Italy}

\abstract{
We extend and refine a general method to extract the multipole moments of arbitrary stationary spacetimes and apply it 
to the study of a large family of regular horizonless solutions to $ {\cal N}{\,=\,}2$ four-dimensional supergravity coupled to four 
Abelian gauge fields. These microstate geometries can carry angular momentum and have a much richer multipolar structure 
than the Kerr black hole. 
In particular they break the axial and equatorial symmetry, giving rise to a large number of nontrivial multipole 
moments.  After studying some analytical examples, we explore the four-dimensional parameter space of this family 
with a statistical analysis. We find that microstate mass and spin multipole moments are typically (but not always) 
larger that those of a Kerr black hole with the same mass and angular momentum. Furthermore, we find numerical 
evidence that some invariants associated with the (dimensionless) moments of these microstates grow monotonically with 
the microstate size and display a global minimum at the black-hole limit, obtained when all centers collide. Our 
analysis is relevant in the context of measurements of the multipole moments of dark compact objects with 
electromagnetic and gravitational-wave probes, and for observational tests to distinguish fuzzballs from classical black 
holes.}
\emailAdd{massimo.bianchi@roma2.infn.it}
\emailAdd{dario.consoli@univie.ac.at}
\emailAdd{alfredo.grillo@roma2.infn.it}
\emailAdd{paolo.pani@uniroma1.it}
\emailAdd{guilherme.raposo@roma1.infn.it}
\begin{document}

\maketitle
\flushbottom

\section{Introduction}

Within classical General Relativity~(GR) a series of 
theorems~\cite{Carter71,Hawking:1973uf,Heusler:1998ua,Chrusciel:2012jk,Robinson} state that the unique vacuum, 
stationary solution is the Kerr metric~\cite{Kerr:1963ud}, which is therefore believed to provide a reliable 
description of the spacetime around {\it any} dark compact object formed after gravitational collapse.

Any stationary Black Hole~(BH) in isolation is axisymmetric. As a result the only non-vanishing mass (current) 
multipole moments are $\mathcal{M}_\ell= \mathcal{M}_{\ell m=0}$ ($\mathcal{S}_\ell= \mathcal{S}_{\ell  m=0}
$), that satisfy the elegant relation~\cite{Geroch:1970cd,Hansen:1974zz} 
\begin{equation}
 \mathcal{M}_\ell+{\rm i }  \mathcal{S}_\ell 
 =\mathcal{M}^{\ell+1}\left({\rm i } \chi\right)^\ell\,, \label{nohair}
\end{equation}
where $\mathcal{M}={\cal M}_0$ is the BH mass, $\mathcal{J}=\mathcal{S}_1$ the angular momentum, and 
$\chi\equiv{\mathcal{J} }/{\mathcal{M}^2}$ the dimensionless spin\footnote{We use $\hbar=c=G_N=1$ throughout.}.  
Equatorial symmetry of the Kerr metric implies ${\cal M}_\ell=0$ 
(${\cal S}_\ell=0$) when $\ell$ is odd (even), and the specific spin dependence of the non-vanishing moments, ${\cal 
M}_\ell\propto\chi^\ell$ and ${\cal S}_\ell\propto\chi^\ell$. This peculiarity of the Kerr metric is not enjoyed by 
other compact-object solutions in GR~\cite{Pani:2015tga,Uchikata:2015yma,Uchikata:2016qku,Raposo:2018xkf}, neither by 
BHs in other gravitational theories~\cite{Psaltis:2008bb,Yunes:2013dva,Berti:2015itd}.

Measuring (at least three) properties of an astrophysical dark object, such as mass, spin, and the mass quadrupole 
$\mathcal{M}_2$, may provide null-hypothesis tests of the Kerr metric and as consequence of Einstein's gravity in the 
strong-field regime~\cite{Psaltis:2008bb,Gair:2012nm,Yunes:2013dva,Berti:2015itd,
Cardoso:2016ryw,Barack:2018yly,Cardoso:2019rvt}. This adds to other observational tests of fuzzballs that have been 
recently proposed, see, e.g., Refs.~\cite{Hertog:2017vod,Guo:2017jmi}.
Quite intriguingly, the current gravitational-wave observations (especially the recent GW190814~\cite{Abbott:2020khf} 
and GW190521~\cite{Abbott:2020tfl,Abbott:2020mjq}) have not yet excluded the 
possible existence of exotic compact objects other than BHs and neutron stars.

According to the cosmic censorship conjecture, curvature singularities in GR are believed to be covered by 
event horizons~\cite{Penrose:1969pc,Wald:1997wa,Penrose_CCC}. A consistent quantum theory of gravity should be able to 
resolve or smoothen BH singularities and to provide a microscopical interpretation of the 
BH thermodynamical properties, such as entropy and temperature,  related to the area of the event horizon and its 
surface gravity, respectively~\cite{Bekenstein,Hawking:1976de}. Furthermore, BH evaporation through the emission of
Hawking radiation~\cite{Hawking:1974sw} leads to other paradoxes, that can be addressed in a  consistent quantum theory 
of gravity such as string theory.
 
In this framework, BHs can be represented as bound-states of strings and D-branes intersecting point-wise along the 
spacetime. Extremal (charged BPS) BHs can be successfully described and a precise microscopic account of the entropy 
can be given through the counting of light excitations of the open strings connecting the various 
branes~\cite{Strominger:1996sh, Horowitz:1996ay, Maldacena:1997de}. 

The information-loss paradox and the singularity problem~\cite{Penrose:1969pc,Wald:1997wa,Penrose_CCC} in GR can be 
solved in string theory relying on the ``fuzzball'' proposal~\cite{Lunin:2001jy, Lunin:2002qf, Mathur:2005zp, 
Mathur:2008nj,Mathur:2009hf}. From this vantage point, BH microstates are associated to smooth horizonless geometries 
with the same asymptotics (mass, charges, and angular momenta). Classical properties of BHs emerge as a result of a 
coarse-graining averaging procedure or as a `collective behavior' of fuzzballs~\cite{Bianchi:2017sds, Bianchi:2018kzy, 
Bena:2018mpb, Bena:2019azk, Bianchi:2020des}.  
Unfortunately, so far, finding a statistically significant fraction 
for five-dimensional (3-charge) and for four-dimensional (4-charge) BPS BHs have proven to be too challenging of a 
task. Only a limited class of microstate geometries have been found, using 
multi-center or stratum ansatze \cite{Bena:2015bea, Bena:2016agb, Bena:2016ypk, Bena:2017xbt, Bianchi:2017bxl, 
Bena:2017upb}, that 
can be embedded in a consistent quantum theory of gravity such as string theory~\cite{Giusto:2009qq, Giusto:2011fy, 
Bianchi:2016bgx}. Very little or nothing is known at the moment about microstates of neutral and non-BPS BHs.

Furthermore, not much has been done to investigate the phenomenological consequences of the fuzzball proposal 
and to identify observables that can distinguish an ensemble of microstates from the classical BH picture or from other 
exotic compact objects which are still viable hypothesis. 
In particular, the measured masses of the binary components of GW190814~\cite{Abbott:2020khf} and of 
GW190521~\cite{Abbott:2020tfl,Abbott:2020mjq} look incompatible with the standard astrophysical formation scenario for 
BHs, being either too light (as in 
the case of the lighest body in GW190814) or too massive (as it seems the case for at least one of the bodies in 
GW190521).
Thus, testing the ``Kerr hypothesis'' is an urgent cornerstone of strong-field verifications of gravity, based on 
different observations with both electromagnetic and gravitational-wave 
probes~\cite{Psaltis:2008bb,Gair:2012nm,Yunes:2013dva,Berti:2015itd, 
Cardoso:2016ryw,Barack:2018yly,Cardoso:2019rvt}.

The scope of this paper (a companion of a recent letter~\cite{Bianchi:2020bxa}) is to study one specific aspect of 
fuzzballs that can be used to distinguish microstates geometries from their classical BH counterpart. Namely, 
we shall study the multipolar structure, which {\it inter alia} affects the motion of test particles around a central 
object, the inspiral of a binary system, and therefore the electromagnetic signal from accreting dark compact 
objects~\cite{Psaltis:2008bb} and the gravitational-wave signal emitted by coalescing binaries~\cite{Blanchet:2006zz}.
Studying the multipolar structure of fuzzballs is particularly interesting for two reasons: 
\begin{itemize}
 \item As argued in~\cite{Bianchi:2020bxa}, the multipolar structure of 
a fuzzball is significantly richer than that of a Kerr BH. While the latter is equatorial and axial symmetric, a 
microstate geometry can generically break any symmetry. This 
results in new classes of multipole moments which are identically zero in the Kerr case~\cite{Raposo:2018xkf}. 
Furthermore, as dictated by the no-hair theorem~\cite{Heusler:1998ua,Chrusciel:2012jk,Robinson}, all properties of a 
Kerr BH --~including of course its infinite tower of multipole moments~-- are determined in terms of its mass ${\cal 
M}$ and 
angular momentum ${\cal J}$. Therefore, measuring independently three arbitrary multipole moments (typically the mass, 
spin, 
and the mass quadrupole moment) can place a strong constraint on alternatives to the classical Kerr 
picture~\cite{Psaltis:2008bb,Gair:2012nm,Yunes:2013dva,Berti:2015itd, 
Cardoso:2016ryw,Barack:2018yly,Raposo:2018xkf,Cardoso:2019rvt}.
 \item At variance with other observables, the multipole moments have the advantage of being easy to calculate, since 
they require only an asymptotic expansion of the metric. This is particularly convenient in the context of fuzzballs, 
since the latter are typically described by very complicated metrics. Furthermore, although microstate geometries are 
manifestly regular when lifted to higher dimensions, they appear singular in four dimensions, the singularity being 
compensated by some divergence of the scalar fields emerging from the sanctification. Although harmless from a 
physical point of view, this singularity (as well as the lack of symmetries) complicates some phenomenological studies, 
for example the computation of the quasi-normal modes of these solutions. On the contrary, the multipole moments are 
extracted at asymptotic infinity, where the solution is manifestly regular also in four spacetime dimensions.
\end{itemize}

In this work we provide full details of the computation presented in Ref.~\cite{Bianchi:2020bxa} and extend that 
analysis to other, more general, solutions.
While our approach is general and applies to any multi-center microstate geometry, we shall focus mostly on 
three-center solutions. As we shall show, the four-dimensional parameter space of this family is very rich. We identify 
some invariants associated with the multipole moments and employ a statistical analysis to compare the multipole 
moments of random microstate geometries with: a)~those 
of a Kerr BH with the same mass and angular momentum; and b)~those of the corresponding solution in the (non-rotating) BH limit, which 
is obtained when all centers collide on a point. In the former case we find that about $90\%$ of the solutions have 
invariant moments larger than Kerr, whereas in the latter case the invariants appear to be always larger than the 
corresponding quantities in the BH limit. Moreover these invariants grow always monotonically with the size (average 
distance between the centers) of the microstate. These properties are analogous to the fact that the quasi-normal mode 
exponential decay rate (the Lyapunov exponent of unstable null geodesics near the photon sphere) is maximum 
for the BH solution~\cite{Bianchi:2020des} and provide a portal to test the fuzzball proposal phenomenologically. 

\section{Multipole moments of generic stationary spacetime}
 
In this section, we introduce two equivalent definitions of the multipole moments which can be directly applied to 
generic stationary and asymptotically flat metrics with no extra symmetry. 

 \subsection{Multipole moments  of the metric}
 We consider stationary asymptotically flat geometries in four dimensions.  
  In an asymptotically Cartesian mass centered (ACMC) system, the metric of a stationary asymptotically flat object can 
be written as~\cite{Bianchi:2020bxa}
\be
\label{eq:ACMC}
ds^2= dt^2 (-1+c_{00})+c_{0i}\, dt \, dx_i +(1+c_{00})\, dx_i^2 +\ldots
\ee
with $c_{00}$ and $c_{0i}$ harmonic functions admitting a harmonic expansion of the 
form 
\begin{equation}
\begin{aligned}\label{cab_cart_coord}
c_{00} &{\,=\,} 2\sum_{\ell=0}^\infty \sum_{m=-\ell}^\ell \frac{1}{r^{1+\ell}}\sqrt{\frac{4\pi}{2\ell{\,+\,} 1}}  
{\cal M}_{\ell m} Y_{\ell m}
\\
c_{0i} &{\,=\,} 2 \sum_{\ell=1}^\infty\sum_{m=-\ell}^{\ell}\frac{1}{r^{1+\ell}}\sqrt{\frac{4\pi (\ell{\,+\,} 1)}{ 
\ell(2\ell{\,+\,} 1)}}   
{\cal S}_{\ell m}Y^{B}_{i;\ell m}
\end{aligned}
\end{equation}
in terms of the scalar ($Y_{lm}$) and axial vector ($Y^{B}_{i;\ell m}$) spherical 
harmonics\footnote{Note that the reality of the metric components, along with the properties of spherical harmonics, imply
the relation ${\cal M}_{\ell,-m}=(-1)^m {\cal M}_{\ell, m}^{\,*}$ (and likewise for the current 
moments).}. In (\ref{eq:ACMC}) and thereafter the dots stand for terms involving spherical harmonics with $\ell'< \ell$ at order $r^{-(1+\ell)}$ in the expansion. 

We use the following definition for the scalar spherical harmonics
\begin{equation}
\label{sph_harm_main}
Y_{\ell m} = \sqrt{\frac{2\ell+1}{4\pi}\frac{(\ell-m)!}{(\ell+m)!}} e^{im \phi} P_{\ell m}(\cos \theta)
\end{equation}
where $P_{\ell m}(x)$ are the associated Legendre polynomials 
 \begin{equation}
P_{\ell m}(x)=\left\{
\begin{aligned}
&\frac{(-)^m(1-x^2)^{m/2}}{2^\ell \ell!}\frac{d^{\ell+m}}{dx^{\ell+m}}(x^2-1)^\ell
\quad
,
\quad
~~~~~m\geq 0
\\
&\frac{(\ell+m)!}{(\ell-m)!}\frac{(1-x^2)^{-m/2}}{2^\ell \ell!}\frac{d^{\ell-m}}{dx^{\ell-m}}(x^2-1)^\ell
\quad
,
\quad
m<0
\end{aligned}
\right.
\end{equation}
For the sake of generality we give here the definition of the (radial, electric, and magnetic) vector spherical 
harmonics\footnote{
Notice that $Y^R_{i;{\ell m}}= {1\over r} X^i, Y_{\ell m}$, $Y^E_{i;{\ell m}} = r P^i Y_{\ell m}$ and $Y^B_{i;{\ell m}} 
= L_i Y_{\ell m}$  where $X^i=x^i$, $P^i = \partial^i$ and $L_i = \varepsilon_{ijk} x^i \partial^k$ are the coordinate, 
momentum, and angular momentum operators, respectively. Moreover only $Y^{B}_{i;\ell m}$ are eigenfunctions of the 
Laplacian $\nabla^2_{S^2}$.  }
\be
\begin{aligned}\label{sph-harm0}
Y^{R}_{i;\ell m}=n_i Y_{\ell m}\,.
\qquad , \qquad
Y^{E}_{i;\ell m} & =\frac{r \partial_{i}Y_{\ell m}}{\sqrt{\ell(\ell+1)}}
\qquad , \qquad
Y^B_{i;\ell m}=\frac{\epsilon_{ij k} \,n_j\,r\partial_{k}Y_{\ell m}}{\sqrt{\ell(\ell+1)}}
\end{aligned}
\end{equation}

The expansion coefficients ${\cal M}_{\ell m}$ and ${\cal S}_{\ell m}$ in Eq.~\eqref{cab_cart_coord} are the mass and 
current multipole moments of the spacetime, respectively. They can be conveniently packed into a single complex 
harmonic 
function defined as
 \begin{align} \label{HH}
H&=H_1+{\rm i}\, H_2= \sum_{\ell=0}^\infty \sum_{m=-\ell}^\ell \frac{1}{r^{1+\ell}}\sqrt{\frac{4\pi}{2\ell+1}}  
\left( {\cal M}_{\ell m} +{\rm i }\,  {\cal S}_{\ell m}  \right) Y_{\ell m}\,.
   \end{align} 
In terms of these variables the ACMC metric (\ref{eq:ACMC}) can be written in the form
\begin{equation}
\begin{aligned}
\label{Asympt_met}
ds^2 &= -e^{-2 H_1} (dt+\omega)^2+ e^{2H_1}  dx_i^2 +\ldots \\
*_3 d\omega &=\epsilon_{ijk}\partial_k c_{0j}dx^i +\ldots   =2 \,dH_2+\ldots \end{aligned}
\end{equation}
with dots standing again for lower harmonics. 

For axi-symmetric solutions (like the Kerr metric) it is convenient to rotate the coordinate axes so that the 
angular momentum vector is aligned with the $z$-axis. In this case the spherical harmonics with $m\neq 0$ 
vanish and one can write (defining from brevity ${\cal M}_{\ell0}\equiv {\cal M}_\ell$ and likewise for the current 
moments)
\begin{align} \label{HH0}
H&=  \sum_{\ell=0}^\infty   \frac{1}{r^{1+\ell}} 
\left( {\cal M}_{\ell } +{\rm i } \, {\cal S}_{\ell }  \right) P_{\ell}(\cos\theta)\,.
\end{align}

\subsection{Multipolar expansion of the Killing one-form associated to stationarity}
The mass (${\cal M}_{\ell m}$)  and spin (${\cal S}_{\ell m}$) multipole moments can be alternatively viewed as the 
``electric'' and ``magnetic''  spherical harmonic expansion coefficients of the Killing one-form $K=   g_{t \mu}dx^\mu$ 
associated to the Killing vector $\partial_t$ of the stationary spacetime. 
 Indeed, inverting formulae (\ref{cab_cart_coord})  one finds 
  \begin{equation}
\begin{aligned}
\mathcal{M}_{\ell m} & = \frac{\sqrt{2\ell+1}}{2(\ell+1)  \sqrt{4\pi} } \,\lim_{r\to\infty}\,r^\ell\int Y_{\ell m}^* 
*dK\\
\mathcal{S}_{\ell m} & = -\frac{\sqrt{2\ell+1}}{2(\ell+1)  \sqrt{4\pi} } \,\lim_{r\to\infty}\,r^\ell\int Y_{\ell m}^* 
dK 
\label{smkill}
\end{aligned}
\end{equation}
  where we used $\nabla^2_{S_2} Y_{\ell m}=-\ell(\ell+1) Y_{\ell m}$.  
 Mass and angular momentum can be read off from the lower multipole moments
\begin{equation}
\mathcal{M} = \mathcal{M}_{00}\,,\qquad |\mathbfcal{J}|  = 
\sqrt{\left|\mathcal{S}_{10}\right|^2+\left|\mathcal{S}_{11}\right|^2+\left|\mathcal{S}_{1-1}\right|^2}\,.
\end{equation}
 
For later convenience, we introduce the dimensionless ratios
\begin{equation}
\overline{\mathcal{M}}_{\ell m} = \frac{\mathcal{M}_{\ell m}}{\mathcal{M}^{\ell+1}}
\quad
,
\quad
\overline{\mathcal{S}}_{\ell m} = \frac{\mathcal{S}_{\ell m}}{\mathcal{M}^{\ell+1}}\,,
\end{equation} 
and the dimensionless spin parameter, $\chi={\cal J}/{\cal M}^2$.

\subsection{Multipolar structure of the Kerr(-Newman) metric}

Although the multipolar structure of the neutral Kerr and charged Kerr-Newman BHs coincide~\cite{Sotiriou:2004ud}, here 
we review the most generic case of the Kerr-Newman solution. 
In the Boyer-Lindquist~(BL) $\{ t, \hat r, \hat \theta,\phi \}$ coordinates the metric and gauge field describing the 
Kerr-Newman solution can be written as\footnote{$t \in (-\infty,+\infty) \,;\, \hat r \in [0,+\infty) \,;\, \hat 
\theta \in [0, \pi] \,;\, \phi \in [0,2\pi]$.}
\begin{align}
ds^2&=-(1{-}\Delta_t) dt^2
-2a \sin^2\theta \Delta_t d t\, d\phi
+\frac{\Sigma}{\Delta_r}d r^2+\Sigma d\hat \theta^2 
+\frac{\sin^2\hat \theta}{\Sigma}\left[( \hat r^2{+}a^2)^2-a^2 \Delta_r \sin^2 \hat \theta\right] d\phi^2 \nn\\
A&=-\frac{Q \hat r}{\Sigma }(dt-a \sin^2 \hat \theta \,d\phi)
-\frac{P \cos \hat \theta}{\Sigma} [a\, dt-(a^2+\hat r^2) \,d\phi]\,,
\end{align}
where
\begin{equation}
\Sigma=\hat r^2+a^2 \cos^2 \hat\theta \quad, \quad
\Delta_t= \frac{2{\cal M} \hat r-(Q^2+P^2)}{\Sigma}  \quad, \quad
\Delta_r= \hat r^2-2{\cal M}\, \hat r+a^2+Q^2+P^2\,.
\end{equation}
This solution is characterised by the mass ${\cal M}$, electric and magnetic charges $Q$ and $P$, and angular momentum 
${\cal J}=a {\cal M}$ defined as 
\begin{equation}
\begin{aligned}
{\cal M} &=\frac{1}{8\pi} \int_{S^2_\infty} 		\!\!*dK \quad,\quad
{\cal J}=-\frac{3}{16\pi} \lim_{r\to \infty} \int_{S^2_r} 	\!\! r \cos \theta\, dK=a M  \\
Q & =\frac{1}{4\pi} \int_{S^2_\infty} 		\!\!*F ~~\quad,\quad
P=\frac{1}{4\pi} \int_{S^2_\infty} 		\!\!F\,.
\end{aligned}
\end{equation}
 Inner and outer horizons exist for masses satisfying ${\cal M}^2\geq Q^2+P^2+a^2$ and are located at
$\hat r_{\pm}={\cal M} \pm \sqrt{{\cal M}^2-a^2-Q^2-P^2}$.  A curvature singularity is found at $\Sigma=0$. The area of 
the BH 
horizon is $A_H=4\pi (\hat r_+^2+a^2)$.

The Kerr-Newman metric in the BL coordinates is not in the ACMC form.  
Indeed, in spherical coordinates a metric in the ACMC form can be written as 
\begin{equation}
ds^2= g_{\mu \nu} dx^\mu dx^\nu =e^a \, e^b (\eta_{ab}+c_{ab})
\qquad
,
\qquad
e^a =  (dt,dr,r d\theta, r \sin \theta d\phi)\,,
\end{equation}
 with $c_{ab}$ such that only harmonics of order at most $\ell$ at order $r^{-\ell+1}$ are present. It is easy to 
see that $c_{\hat r \hat r}$ and $c_{\hat \theta \hat \theta}$ in the Kerr-Newman metric in BL coordinates fail to meet 
this requirement. To bring the metric to the ACMC form one can perform the coordinate transformation  
 \begin{equation}
r ^2  = ( \hat r-{\cal M})^2+a^2 \sin^2 \hat \theta \qquad , \qquad   r\cos\theta = (\hat r-{\cal M})\cos\hat \theta\,,
\end{equation}
which reduces to the (perturbative) result found by Hartle and Thorne~\cite{Hartle:1968si} to second order in the spin.

In the new variables one finds the non-vanishing components
 \begin{align}\label{ckerr}
 c_{00} &=c_{rr}=c_{\theta\theta}=c_{\phi\phi}={2{\cal M}\over r} \sum_{n=0}^\infty (-)^n {a^{2n}\over r^{2n} }  
P_{2n}(\cos\theta)+\ldots \nn\\
 c_{0\phi} &= {2 a {\cal M} \over r^2} \sum_{n=0}^\infty    (-)^n {a^{2n}\over r^{2n} }  {\partial_\theta  
P_{2n-1}(\cos\theta)  \over 2n-1}   +\ldots 
\end{align}
 leading to~\cite{Hansen:1974zz,Geroch:1970cd}
\be
\mathcal{M}_{2n}= (-1)^n a^{2n} {\cal M} \qquad , \qquad \mathcal{S}_{2n+1}= (-1)^n a^{2n+1} {\cal M}. \label{kerrm}
\ee
 The mass and current multipole moments combine into the single complex harmonic function
\be
H_{\rm Kerr}= \sum_{\ell=0}^\infty  \left(\mathcal{M}_{\ell}+{\rm i}\, \mathcal{S}_{\ell} \right) \,   \frac{  P_{\ell 
} 
 }{r^{1+\ell}} = \frac{{\cal M}}{\sqrt{ x_1^2+x_2^2+(x_3-{\rm i} a)^2 } }
\ee
We notice that real and imaginary parts of $H_{\rm Kerr}$ are given in terms of a sort of analytic continuation of a 
two-center harmonic function with centers located at $\pm i a$. In particular the mass is ${\cal M}$ and the 
angular momentum ${\cal J}=a{\cal M}$. 
The Schwarzschild solution is obtained by sending $a\to 0$ and the two centers coincide at the origin. 

Finally, note that the multipole moments of the Kerr-Newman metric, Eq.~\eqref{kerrm}, do not depend explicitly on the 
charges, so they are the same in the neutral (Kerr) limit~\cite{Sotiriou:2004ud}. Obviously the same holds true 
for Reissner-Nordstr\"om (the $\chi\to0$ limit of Kerr-Newman), whose multipolar structure is the same as for 
Schwarzschild.  More generally, the presence of minimally coupled 
scalar and gauge fields, equipped with energy momentum tensors dying faster than $1/r^3$ at infinity, does not
destroy the Ricci flatness of the leading harmonic part of the ACMC metric (and hence it does not affect the way 
multipole moments can be extracted), but nevertheless results in a very different (from Kerr) multipolar structure.

\section{Fuzzball solutions and their multipolar structure}
Fuzzball solutions can be viewed as multi-center generalizations of the single center Schwarzschild and two-center Kerr 
metrics. In this section we review a family of solutions and discuss their multipolar structure.

\subsection{The metric}
In the framework of ${\cal N}=2$ four-dimensional supergravity, we consider gravity minimally coupled 
to four Maxwell fields and three complex scalars. A general class of extremal solutions of the Maxwell-Einstein-scalar 
system is described by a  metric of the form~\cite{Bena:2007kg,Gibbons:2013tqa,Bates:2003vx} 
\begin{align} \label{4dsolution}
ds^2=-e^{2U}\pt{dt+\omega}^2+e^{-2U}\sum_{i=1}^3 dx^2_i, 
   \end{align}  
with
\be
e^{-4U}{=}VL_1L_2L_3{-}K^1K^2K^3M{+}\ft12\sum_{I>J}^3K^IK^JL_IL_J  {-}{MV\over 2}\sum_{I=1}^3K^IL_I{-}{M^2V^2\over 4}
{-}\ft14 \sum_{I=1}^3 (K^I L_I)^2
 \label{expmin4U}
\ee
\be
 *_3d\omega = \half\pt{VdM-MdV+K_IdL_I-L_IdK_I} 
\label{dStarOmega}
\ee
 and $ \{ V, L_I, K^I, M \} $ eight harmonic functions,  $I=1,2,3$. We consider $N$-center harmonic functions  
 \bea
 V &=& v_0+\sum_{a=1}^N  {v_a\over r_a }  \qquad  ,  \qquad      L_I = {\ell}_{0I}+ \sum_{a=1}^N   { {\ell}_{I,a} \over  
r_a }  \nn\\ 
 K^I &=& k^I_{0}+\sum_{a=1}^N  {k^I_{a}\over r_a }  \qquad  ,  \qquad M =  m_0+\sum_{a=1}^N   { m_a \over  r_a }  
\label{ansatz0}
 \eea
 with $ r_a =|{\bf x}-{\bf  x}_a|$  and  ${\bf x}_a$ the position of the $a^{\rm th}$ center. The quantities 
$({\ell}_{Ia} , m_a)$ and $(v_a,k^I_a)$ describe the electric and magnetic  fluxes of the four-dimensional gauge fields, 
so Dirac quantisation requires that they be quantised. Here we adopt units such that they are all integers.

 Charges and positions of the centers can be chosen such that the metric near the center lifts to a smooth 
five-dimensional geometry of type $\mathbb{R}_t$ times a Gibbons-Hawking space. This requirement boils down to a 
restriction on the $k_a^J$ known as bubble equations \cite{Bena:2007kg}

\begin{equation}
\label{bubble-general}
\sum_{b=1}^N\Pi_{ab}^{(1)}\Pi_{ab}^{(2)}\Pi_{ab}^{(3)}\frac{v_a v_b}{r_{ab}} + v_0 \frac{k_a^1 k_a^2 k_a^3}{v_a^2} - \sum_{I=1}^3\ell_{0I} k_a^I - \left|\epsilon_{IJK}\right| \frac{k_{0}^Ik_a^J k_a^K}{2v_a} - m_0 v_a = 0
\end{equation}
where
\begin{equation}
\Pi_{ab}^{(I)} = \frac{k_a^I}{v_a} - \frac{k_b^I}{v_b}
\end{equation}
and to the following relations
\begin{equation}
\ell_{I,a}=-{1\over 2} |\epsilon_{IJK}| k_a^J k_a^K  
\quad 
,
\quad 
m_a =k_a^1 k_a^2 k_a^3
\end{equation}

On the other hand, at infinity the five-dimensional geometry looks like a four-charged BH in four dimensions times a circle.
The parameters $k^I_a$ and the positions of the centers describe the charges and moduli of the microstate.

Furthermore the regularity conditions
 \be
 e^{2U} >0 \qquad , \qquad L_I V+\frac{1}{2} |\ve_{IJK}| K^JK^K>0\,, \label{regularitycond}
 \ee
ensuring the absence of horizons and of closed time-like curves, should be imposed. 

Finally, even though our derivation of the multipole moments in the following section is completely general, we will later focus on fuzzballs of intersecting orthogonal branes (considered in \cite{Bianchi:2017bxl}) for which the following conditions hold
\begin{equation}
\label{asympt_cond_general}
\ell_{0I} = v_0 = v_a = 1
\quad
,
\quad
m_0 = k_{0}^I = 0
\end{equation}

\subsection{The multipole moments}

 The metric (\ref{4dsolution}) is already in the ACMC form. Restricting to leading harmonic components, the 
formulae~\eqref{expmin4U},\eqref{dStarOmega} lead to
  \bea
 e^{-4U}  & =& 1+4 \sum_{a=1}^N {\mu_a \over r_a} +\ldots \nn\\
  *_3d\omega &=& \half  d (v_0 \, M-m_0 \, V+k^I_0 L_I-\ell_{0,I} K^I) +\ldots = 2  d  \left( \sum_{a=1}^N { j_a \over r_a } \right)  
+\ldots 
 \eea
where $\mu_a$ and $j_a$ are some rational numbers following from the expansion of the left-hand side and  we discard 
terms dying faster than $r_a^{-1}$ in the limit of large $r_a$ since they contribute to lower harmonic components. 
Comparing with (\ref{Asympt_met}), one finds that the complex harmonic function $H=H_1+{\rm i} H_2$ can then be written 
as a sum over centers
 \be
 H =\sum_{\ell=0}^\infty \sum_{m=-\ell}^\ell \sqrt{\frac{4\pi}{2\ell+1}}  \frac{Y_{\ell m} 
}{r^{1+\ell}}(\mathcal{M}_{\ell m} +{\rm i} \mathcal{S}_{\ell m} )  
 =  \sum_{a=1}^N { \mu_a+ {\rm i} j_a \over r_a}  +\dots
 \ee  
Using the harmonic expansion
  \begin{equation}
\label{1_ra_exp}
\begin{aligned}
\frac{1}{r_a}  = \sum_{\ell=0}^\infty \sum_{m=-\ell}^\ell  \sqrt{\frac{4\pi}{2\ell+1}} \frac{R_{\ell m}^a Y_{\ell 
m}(\theta,\phi)}{r^{\ell+1}}
\end{aligned}
\end{equation}
 with
\begin{equation}
R_{\ell m}^a = |{\bf x}_a|^\ell \sqrt{\frac{4\pi}{2\ell+1}} \, Y_{\ell m}^*(\theta_a,\phi_a)	
\end{equation}
 one finds the compact result
\begin{equation}
\label{general_multipoles}
\begin{aligned}
\mathcal{M}_{\ell m} + {\rm i} \mathcal{S}_{\ell m} &=  \sum_{a=1}^N (\mu_a+{\rm i} j_a) R_{\ell m}^a\,.
\end{aligned}
\end{equation}
We put the origin of the coordinate system in the center-of-mass and orient the $z$-axis along the angular momentum, so  
that
\begin{equation}
 \begin{aligned}
 \sum_{a=1}^N  \mu_a {\bf x}_a  =  0\quad,\quad
  \mathbfcal{J}=\sum_{a=1}^N  j_a {\bf x}_a  =  {\cal J} {\bf e}_z
\end{aligned}
\end{equation}
with ${\bf e}_z$ the unit vector along $z$. With this choice ${\cal M}_{1m}=0$, ${\cal S}_{1\,\pm 1}=0$, 
and ${\cal S}_{10}={\cal J}$.

\subsection{Single-center solutions}

Single center solutions correspond to Reissner-Nordstr{\"o}m BHs, their multipole structure coincides with the one of 
Schwarzschild BHs and reads 
\begin{equation}
\mathcal{M}_{00} = \mathcal{M}\,,\quad \mathcal{M}_{\ell>0, m} = 0\,,\quad  \mathcal{S}_{\ell m} = 0\,.
\end{equation}

\subsection{Two-center solutions}

With only two centers, under the assumptions (\ref{asympt_cond_general}), there is no solution to the bubble equations~\eqref{bubble-general}
\footnote{Two-center regular geometries with asymptotics differing from  (\ref{asympt_cond_general}) were obtained in \cite{Denef_2011,Raeymaekers_2008,Boer_2008}.}. Still, it is interesting to observe that the multipole moments of singular two-center solutions bear some similarity with those recently obtained in~\cite{Bena:2020see} for non-extremal STU BHs, with some important differences. One can always align the center position along the $z$-axis, \textit{i.e.} $\theta_1 = 0$, 
$\theta_2 = \pi$ and, by requiring that the center of mass lies at the origin, we get
\begin{equation}
\textbf{x}_{1} = \frac{\mu_2}{\mathcal{M}}\,L\,\textbf{e}_z\,,\quad \textbf{x}_{2} = -\frac{\mu_1}{\mathcal{M}} L \, 
\textbf{e}_z\,.
\end{equation}
The multipole moments follow then from (\ref{general_multipoles}) and read\footnote{The multipole moments of a non extremal STU BHs are given by 
(following the notation of Ref.~\cite{Bena:2020see})
\begin{equation}
\begin{aligned}
{\cal M}_\ell =-\frac{{\rm i} }{2}\left(-\frac{a}{{\cal M}}\right)^\ell 
Z\overline{Z}\left(Z^{\ell-1}-\overline{Z}^{\ell-1}\right)\,,\quad
{\cal S}_\ell 
=\frac{{\rm i} }{2}\left(-\frac{a}{{\cal M}}\right)^{\ell-1}\frac{{\cal J}}{{\cal M}}\left(Z^{\ell}-\overline{Z}^{\ell}\right)
\end{aligned}
\end{equation}
with $Z = D - {\rm i} {\cal M}$.  We notice that the multipole structure of the STU BH depends on four independent 
parameters: ${\cal M}$, ${\cal J}$, $a$, $D$.}
\begin{equation}
\mathcal{M}_{\ell0} = \left(-\frac{L}{\mathcal{M}}\right)^\ell \left[\mu_1^{\ell} \mu_2{\,-\,}({\,-\,}\mu_2)^{\ell} 
\mu_1\right] \quad, \quad
\mathcal{S}_{\ell0} = \left(-\frac{L}{\mathcal{M}}\right)^{\ell} \left[\mu_1^\ell\, j_2{\,+\,}({\,-\,}\mu_2)^\ell\, 
j_1\right]\,.
\end{equation}
Notice that these solutions depend on five parameters, $\mu_1$, $\mu_2$, $j_1$, $j_2$, and $L/\mathcal{M}$.
The multipole expansion of the two-center solution significantly simplifies for $\mu_1 = \mu_2 = \mathcal{M}/2$, $j_1 = 
-j_2 = j$. The non-trivial moments in this case are
\begin{equation}
\begin{aligned}
\mathcal{M}_{2n\,0} = \frac{\mathcal{M}L^{2n}}{2^{2n}}\,,\qquad
\mathcal{S}_{2n+1\,0} = -\frac{j L^{2n+1}}{2^{2n}}\,, \qquad
n \geq 0
\end{aligned}
\end{equation}
and they differ from those of Kerr BHs, Eq.~\eqref{kerrm}, only on the missing alternating sign\footnote{Interestingly enough, they formally match if one chooses $L = 2ia$ and $j = 
i\mathcal{M}/2$.}.

 \subsection{Three-center solutions}
 
Solutions of the bubble equations exist for $N\geq 3$.  For concreteness, here we focus on a simple class of three 
center solutions defined by taking
  \be 
  \begin{aligned}
  v_i\,&{\,=\,}1 \qquad , \qquad \ell_{I,a}=-{1\over 2} |\epsilon_{IJK}| k_a^J k_a^K  \qquad , \qquad m_a =k_a^1 k_a^2 k_a^3
 \\v_0\,&{\,=\,}1 \qquad , \qquad m_0=0  \qquad , \qquad \ell_{0I}=1  \qquad , \qquad k_{0}^I=0
  \label{asympt_cond}
  \end{aligned}
  \ee
with $k_a^J$ satisfying the bubble equations
  \be
  \sum_{b\neq a}  { 1\over r_{a,b} }\prod_{I=1}^3 (k_a^I-k_j^I) +k_a^1 k_a^2 k_a^3 -\sum_{I=1}^3 k_a^I =0\,.  
\label{bubble!!!}
  \ee
 In addition, one has to impose  the regularity conditions (\ref{regularitycond}). 
 For this choice one finds
\begin{equation}
\begin{aligned}
H_1 & =  {1\over 4}\left(V+\sum_{I=1}^3 L_I -4 \right) +\ldots =\sum_{a=1}^3  {\mu_a\over r_a}  +\ldots\\
H_2 & =\frac{1}{4}\left(M - \sum_{I=1}^3 K^I\right) +\ldots= \sum_{a=1}^3 {j_a\over r_a} +\ldots
\label{dw3}
\end{aligned}
\end{equation}
where dots again refer to lower harmonic contributions and
\begin{equation}
\mu_a = \frac{1}{4}\left(v_a+\sum_{I=1}^3\ell_{I,a}\right)\,,\quad j_a = \frac{1}{4}\left(m_a-\sum_{I=1}^3 k_a^I\right)\,.
\end{equation}
 Multipole moments are then given by  
\begin{equation}
\label{general_multipoles3center}
\begin{aligned}
\mathcal{M}_{\ell m} + {\rm i} \mathcal{S}_{\ell m} &=  \sum_{a=1}^3 (\mu_a+{\rm i} j_a) R_{\ell m}^a\,.
\end{aligned}
\end{equation}
 Solutions will be labeled by four integers $\vec\kappa=(\kappa_1,\kappa_2,\kappa_3,\kappa_4)$ and a length scale $L$. 
We will consider various limits where some $\kappa$'s and/or $L$ become large or small. 
  We will often display formulae for the following dimensionless ratios
  \be
\mathfrak{M}_{\ell 0} = { \overline{\cal M}_{\ell 0} \over \chi^{\ell} }  \qquad , \qquad \mathfrak{S}_{\ell 0} = { 
\overline{\cal S}_{\ell\, 0} \over \chi^{\ell} }\,, \label{mathfrak}
  \ee
where $\ell$ is even and odd, respectively.
These ratios are $\pm 1$ for Kerr BHs and are ill-defined for Schwarzschild BHs (see \cite{Yagi:2016bkt} and references 
therein for early studies of these ratios in the context of compact objects within GR and beyond). 
Furthermore, in the case of neutron stars~\cite{Hartle:1967he,Hartle:1968si,Yagi:2016bkt} and boson 
stars~\cite{Ryan:1996nk} they are always larger than in the BH case, although these solutions are not continuously 
connected to the BH solution. For other exotic compact objects that continuously connect to the BH metric 
(\textit{e.g.}, gravastars or strongly anisotropic stars) these ratios approach the Kerr value in the BH 
limit~\cite{Yagi:2015upa,Pani:2015tga,Uchikata:2015yma,Uchikata:2016qku,Raposo:2018xkf}. In the context of microstates 
these ratios have been recently studied in~\cite{Bena:2020see,Bena:2020uup}.

We will provide some evidence that mass and current multipole moments of 
microstate solutions are typically (but not always) bigger than those of a Kerr BH with the same 
mass and angular momentum. 
Furthermore, it is convenient to define the quadratic invariants which are proportional to
\begin{eqnarray}
 \mathfrak{M}_\ell \equiv \sqrt{\sum_{m=-\ell}^\ell | \mathfrak{M}_{lm}|^2} \,,\qquad\qquad
\mathfrak{S}_\ell \equiv \sqrt{\sum_{m=-\ell}^\ell |\mathfrak{S}_{lm}|^2}\,.  \label{invariants}
\end{eqnarray}
More general invariants can be built analogously (see Appendix~\ref{app:invariants} for details). Note that the above 
relations reduce to the standard definitions of ${\cal M}_\ell$ and ${\cal S}_\ell$ in the axisymmetric case, modulo 
the sign. We will provide numerical evidence that for three-center microstate geometries these invariants 
grow monotonically with the size $L$ of the microstate, with a global minimum at $L=0$, where the microstate reduce to 
a spherical BH.

\subsubsection{The metric} 
 
The simplest, regular, horizonless geometries arise for three-center solutions. We restrict ourselves to fuzzballs of 
four-charged BHs obtained from orthogonal branes, so we require that $K^I$ and $M$ vanish at order $1/r$, \textit{i.e.}
\be
\sum_{a=1}^3 k_a^I =\sum_{a=1}^3 k_a^1 k_a^2 k_a^3 =0\,.
\ee
 These conditions determine the $k_a^I$ to be of the form
 \be
 k^I{}_a = \left(
\begin{array}{ccc}
 -\kappa_1 \kappa_2 &  -\kappa_1 \kappa_3  &  \kappa_1(\kappa_2+\kappa_3) \\
  \kappa_3 &    \kappa_2  &  -\kappa_2-\kappa_3 \\
   -\kappa_4  &  \kappa_4  & 0 \\  
\end{array}
\right).  
 \ee
 and therefore
 \be
\begin{aligned}
V &= 1+\sum_{a=1}^3 {1\over r_a}\,,   \quad   M=  \kappa_1 \,\kappa_2 \,\kappa_3\, \kappa_4  \left( { 1\over 
r_1}-{1\over r_2}\right)
\\
 L_1 &=1+  \kappa_4 \left( {\kappa_3 \over r_1}-{\kappa_2\over r_2} \right)\,,  \quad  L_2 =1+  \kappa_1 \kappa_4 
\left( 
 -{\kappa_2\over r_1}+{\kappa_3\over r_2} \right)
\\
 L_3 &=1+  \kappa_1 \left( {\kappa_2\kappa_3 \over r_1}+{\kappa_2 \kappa_3\over r_2} +{(\kappa_2+\kappa_3)^2 \over r_3} 
\right)\,, \quad
K^1 =  \kappa_1 \left( -{\kappa_2 \over r_1}-{\kappa_3\over r_2}+{\kappa_2+\kappa_3\over r_3} \right)\\
  K^2 &=  {\kappa_3 \over r_1}+{\kappa_2 \over r_2}-{\kappa_2+\kappa_3 \over r_3}\,,  \quad 
  K^3 =  \kappa_4 \left( -{1\over r_1}+{1\over r_2} \right)
\end{aligned}
 \ee
 with $r_a =| {\bf x}-{\bf x}_a|$ and $\kappa_i$ some arbitrary integers. 
Finally, the bubble equations  (\ref{bubble!!!}) constrain the distances $r_{ab}=| {\bf x}_a-{\bf x}_b|$ between the 
centers to be related by
 \bea
r_{12}&=& 
\frac{2\kappa_1\kappa_4(\kappa_2-\kappa_3)^2r_{23}}{
\kappa_1\kappa_4(2\kappa_2^2+5\kappa_2\kappa_3+2\kappa_3^2)+(\kappa_2+\kappa_4-\kappa_1\kappa_3+\kappa_1\kappa_2\kappa_3
\kappa_4)r_{23}}\nn\\
r_{13}&=& 
\frac{\kappa_1\kappa_4(2\kappa_2+\kappa_3)(\kappa_2+2\kappa_3)r_{23}}{
\kappa_1\kappa_4(2\kappa_2^2+5\kappa_2\kappa_3+2\kappa_3^2)-(\kappa_1-1)(\kappa_2+\kappa_3)r_{23}}.
\label{bubble2!!!}
\eea
The solution describes a microstate of a Reissner-Nordstr\"om BH with a magnetic charge $P_0$ and three electric charges 
$Q_I$ given by 
\be
P_0 {\,=\,} 3 \quad ,\quad 
Q_1 {\,=\,} \kappa_4(\kappa_3{\,-\,}\kappa_2)\quad ,\quad
Q_2 {\,=\,} \kappa_1\kappa_4(\kappa_3 {\,-\,}\kappa_2) \quad ,\quad  
Q_3 {\,=\,} \kappa_1(\kappa_2^2{\,+\,}4\kappa_2\kappa_3{\,+\,}\kappa_3^2) \,.\nn\\
\ee
Besides the parameters $\kappa_i$ describing the BH charges, the solution is described by a continuous 
parameter $r_{23}$ labelling the microstate. 
To have non-zero and positive charges\footnote{For BPS-ness it is enough that the charges be of the same sign and ${\cal 
I}_4(P_0, Q_I, P^I, Q_0)>0$.}  we require
\begin{equation}
\kappa_1 > 0 \quad, \quad \kappa_4 >0 \quad, \quad \kappa_3 > \kappa_2\geq 0\,.
\label{general_kappa}
\end{equation}  
One can check that for $\kappa_3>\kappa_2$ the regularity conditions~\eqref{regularitycond} are always 
satisfied, so from now on $\kappa_3>\kappa_2>0$ will be always assumed.

Finally, the mass and angular momentum of the solution are given by  
 \bea
 {\cal M}=\mu_1+\mu_2+\mu_3
 \quad, \quad
 \mathbfcal{J}= \sum_{a=1}^3 j_a {\bf x}_a = {\cal J} \textbf{e}_z
\label{mass_and_j}
\eea
with ${\bf x}_a$ the positions of the centers, and
\begin{equation}
\begin{aligned}\label{mujs}
 \mu_1 &= \frac{1}{4}(1+\kappa_1\kappa_2\kappa_3-\kappa_1\kappa_2\kappa_4+\kappa_3\kappa_4)
\,,\quad
\mu_2 = \frac{1}{4}(1+\kappa_1\kappa_2\kappa_3+\kappa_1\kappa_3\kappa_4-\kappa_2\kappa_4)
\,,
\\
\mu_3 &= \frac{1}{4}(1+\kappa_1(\kappa_2+\kappa_3)^2)
\,,
\\
j_1 &= \frac{1}{4}[\kappa_2+\kappa_4+\kappa_1\kappa_3(\kappa_2\kappa_4-1)+(\kappa_1-1)(\kappa_2+\kappa_3)]
\\
j_2 &= -\frac{1}{4}(\kappa_2+\kappa_4+\kappa_1\kappa_3(\kappa_2\kappa_4-1))
\,,
\quad
j_3 = -\frac{1}{4}(\kappa_1-1)(\kappa_2+\kappa_3)\,.
\end{aligned}
\end{equation}
Notice that $j_1{\,+\,}j_2{\,+\,}j_3=0$, so much so that $\mathbfcal{J}$ is invariant under rigid translations of the 
centers. 
  
\subsubsection{The location of the centers}

We define our coordinate system such that the three vertices lie on the $(x,z)$-plane ($y_a=0$), with the center of mass 
at the origin and the angular momentum aligned along the (positive) $z$ direction. So we take
\be
{\bf x}_a =(x_a, 0, z_a) 
\ee  
with
\bea
x_a &=& \alpha\, \epsilon_{a b c} \, \mu_b\, j_c\,,\quad z_1 = \beta \mu_2 +\gamma \mu_3 \quad , \quad z_2 = -\beta 
\mu_1  \quad , \quad z_3 = -\gamma \mu_1 \,, \label{xzsol}
\eea
 and $\alpha$, $\beta$, $\gamma$ three parameters to be determined. It is easy to see that this choice satisfies the 
defining conditions 
 \be
 \sum_{a=1}^3 \mu_a x_a= \sum_{a=1}^3 \mu_a z_a = 0 = \sum_{a=1}^3 j_a x_a \,.
 \ee
 The parameters $\alpha$, $\beta$, $\gamma$ are determined by the bubble equations that yield
 \begin{equation}\label{bubcen}
r_{23}=L \qquad, \qquad r_{13}=\sqrt{\rho} L  \quad, \quad
r_{12}= \sqrt{\sigma} L  
\end{equation}
 with
 \be
 \sqrt{\rho} = \frac{1}{1-\tilde{a}_1 L}  \qquad , \qquad    \sqrt{\sigma}= \frac{ \tilde{a}_2}{1+\tilde{a}_3 L}
 \ee
 and
\bea
\tilde{a}_1 & =& 
\frac{(\kappa_1{-}1)(\kappa_2{+}\kappa_3)}{\kappa_1\kappa_4(2\kappa_2{+}\kappa_3)(\kappa_2{+}2\kappa_3)} ~,~ 
\tilde{a}_2 = \frac{2(\kappa_3{-}\kappa_2)^2}{(2\kappa_2{+}\kappa_3)(\kappa_2{+}2\kappa_3)}  ~,~ \label{tildea}
\\
\tilde{a}_3 & =& 
\frac{\kappa_2{+}\kappa_4{+}\kappa_1\kappa_3(\kappa_2\kappa_4{-}1)}{\kappa_1\kappa_4(2\kappa_2{+}\kappa_3)(\kappa_2{+}
2\kappa_3)}\,. \nn
\eea
Under the assumptions \eqref{general_kappa}, we find that the parameters 
$\tilde{a}_i$ always span a finite domain 
 \be
0\leq \tilde a_1 \leq {1\over 2} \quad, \quad 0\leq \tilde a_2 \leq 1 \quad, \quad  -{1 \over 2} \leq \tilde a_3 \leq 
{1\over 2}\,.
 \ee
Solving the bubble equations one finds
\bea
\label{ruleabc}
\alpha&=& {L\sqrt{2(\rho{\,+\,}\sigma){\,-\,}1{\,-\,}(\rho{\,-\,}\sigma)^2} \over 2 {\cal M} \sqrt{j_2^2+ \rho 
j_1^2+(1-\sigma+\rho)j_1 j_2}} \nn\\
\beta &=& {L \left[j_1 (\mu_1 (\rho +\sigma -1)-\mu_3 (1+\rho-\sigma))-j_2 (\mu_1 (1-\rho +\sigma )+2 \mu_3)  \right]
\over 2 \mu_1  {\cal M} \sqrt{j_2^2+ \rho j_1^2+(1-\sigma+\rho)j_1 j_2}} \nn\\
\gamma &=&{L \left[ j_1 (\mu_2 (1+\rho -\sigma)+2 \mu_1 \rho )+j_2 (\mu_1 (1+\rho -\sigma )+2 \mu_2) \right]
  \over 2 \mu_1  {\cal M} \sqrt{j_2^2+ \rho j_1^2+(1-\sigma+\rho)j_1 j_2}}  
 \eea
Solutions exist only if the argument of the square root in the numerator of $\alpha$ is positive.\footnote{One can 
check that the argument of the square root in the denominators is always positive.} Together
with the positivity of $r_{ab}$, one finds that solutions exist for $0<L<L_{\rm max}$ with 
   \be
L_{\rm max} = \frac{(1+\tilde{a}_2)}{2 \tilde{a}_3}\left(\sqrt{1+\frac{4 \tilde{a}_2 
\tilde{a}_3}{\tilde{a}_1 (\tilde{a}_2+1)^2}}-1\right)\,,
\label{Lmax}
 \ee
obtained by carefully inspecting the following inequalities
  \be
\tilde{a}_3 L  \geq - 1 \quad, \quad 0\leq \tilde{a}_1 L \leq 1 \quad, \quad
2(\rho{\,+\,}\sigma){\,-\,}1{\,-\,}(\rho{\,-\,}\sigma)^2\geq 0\,.
 \label{ineq}
 \ee
For $L=L_{\rm max}$ (\textit{i.e.} at the boundary of the third inequality) the parameter $\alpha$ vanishes and the centers are aligned along the $z$-axis, therefore the solution is axisymmetric. Since $\rho$ and $\sigma$ respectively diverge when the second and first inequality above are saturated on the right, it is easy to see that the last inequality is often the most stringent one.

We can distinguish two main classes of solutions:  
   
\begin{itemize}
   
   \item{ $\kappa_1=1$: For this choice $\tilde a_1=0$, $\rho=1$ so the triangle formed by the 
three centers is isoscele or equilateral. The conditions~\eqref{ineq} are always satisfied so solutions exist for any 
choice of $L$. 
    } 
    \item{ $\kappa_1 \neq 1$: This is the generic case, solutions exist only inside the finite 
domain $L\in [0,L_{\max} ]$.
 } 
 \end{itemize}

\subsubsection{The multipole moments}

The resulting expressions for the multipole moments read 
\begin{equation}
\label{general_multipoles2}
\begin{aligned}
\mathcal{M}_{\ell m} + {\rm i} \mathcal{S}_{\ell m} &=  \sum_{a=1}^N (\mu_a+{\rm i} j_a) R_{\ell m}^a\,.
\end{aligned}
\end{equation}
with
\begin{equation}
R_{\ell m}^a = |{\bf x}_a|^\ell \sqrt{\frac{4\pi}{2\ell+1}} \, Y_{\ell m}^*(\theta_a,\phi_a)	
\end{equation}
  and
  \bea
{\bf x}_1 &=&  \left( \alpha\, (\mu_2\, j_3{-}\mu_3 j_2), 0, \beta \mu_2 {+}\gamma \mu_3 \right) \qquad , \qquad  
{\bf x}_2 = \left( \alpha\, (\mu_3\, j_1{-}\mu_1 j_3),0,  {-}\beta \mu_1 \right) \nn\\
{\bf x}_3 &=&  \left( \alpha\, (\mu_1\, j_2{-}\mu_2 j_1),0,  {-}\gamma \mu_1 \right)  
\eea
so that
\be
\cos\theta_a={\alpha\,\epsilon_{abc}\,\mu_b j_c \over |{\bf x}_a| } \qquad , \qquad   \cos\phi_a={\rm sign}  \left( 
\epsilon_{abc}\,\mu_b\, j_c\right)
\ee
   The parameters  $\alpha$, $\beta$, $\gamma$ are given by (\ref{ruleabc}) while $\mu_a,j_a$ are listed in 
(\ref{mujs}). 
  The mass and angular momentum of the solution are given by  
 \bea
 {\cal M}=\mu_1+\mu_2+\mu_3
 \quad, \quad
{\cal J} =  j_1 ( \beta \mu_2 +\gamma \mu_3+ \gamma \mu_1)+ j_2 (\gamma-\beta) \mu_1 \,.
\label{mass_and_j_bis}
\eea
     
\subsection{Examples of three-center solutions}

In this section we present the multipole moments for several interesting examples of the three-center 
family of solutions. The general cases are presented in Appendix~\ref{app:general}. 
 
\subsubsection{Solution~${\bf A}$: $\vec\kappa=(1,0,\lambda,\lambda)$, scaling solution}

The scaling solution is characterized by the following choice of the parameters:
\begin{align}
\kappa_i=&\,(1,0,\lambda, \lambda)
\quad, \quad
\mu_a=\left(\frac{1{\,+\,}\lambda^2}{4},\frac{1{\,+\,}\lambda^2}{4},\frac{1{\,+\,}\lambda^2}{4}\right)
\quad, \quad
j_a=(0,0,0)\\
P_0=&\,3
\quad, \quad 
Q_I=(\lambda^2,\lambda^2,\lambda^2)
\quad, \quad 
{\cal M} = \frac{3(1{\,+\,}\lambda^2)}{4}
\quad, \quad 
{\cal J}=0 \\
\tilde{a}_i=\,&\left( 0, 1, 0\right)
\quad, \quad 
\rho = \sigma = 1
\,.
\end{align}
Therefore $r_{12}{\,=\,}r_{23}{\,=\,}r_{13}{\,=\,}L$, implying that the three centers 
are the vertices of an equilateral triangle. 
Since $\tilde{a}_1 = 0$ the parameter $L$ is unbounded.

The non-trivial mass multipole moments are\footnote{The expression for the mass multipoles is different from the 
correspondent one in \cite{Bianchi:2020bxa} since, at variance with what we do here, in \cite{Bianchi:2020bxa} vertices 
were taken to be lying on the $(x,y)$-plane.} 

\begin{equation}
\mathcal{M}_{\ell\,,2p-\ell} = 
 \frac{\sqrt{(2\ell-2 p)!}}{\sqrt{(2p)!}}\mathcal{M}  \left(-\frac{L}{\sqrt{3}}\right)^\ell\left[P_{\ell\,,2p-\ell}(0)+2(-1)^\ell 
P_{\ell\,,2p-\ell}\left(\frac{\sqrt{3}}{2}\right)\right]
\end{equation}
with $p = 0,1,\ldots,\ell$, while all
current multipoles vanish: $\mathcal{S}_{\ell m} = 0$. 
More specifically, the first nonvanishing moments are 
\begin{equation}
\overline{\mathcal{M}}_{2,0}=\frac{L^2}{4{\cal M}^2}
\quad
,
\quad
\overline{\mathcal{M}}_{2,2}=\frac{1}{4}\sqrt{\frac{3}{2}}\frac{L^2}{{\cal M}^2}
\quad
,
\quad
\overline{\mathcal{M}}_{3,1}= -\frac{5}{16}\frac{L^3}{{\cal M}^3}
\quad
,
\quad
\overline{\mathcal{M}}_{3,3}= \frac{1}{16}\sqrt{\frac{5}{3}}\frac{L^3}{{\cal M}^3}\,,
\end{equation}
and so on. The moments with $m<0$ are given by ${\cal M}_{\ell,-m}=(-1)^m {\cal M}_{\ell, m}^{\,*}$.

Unlike the Kerr case, the mass quadrupole is non-vanishing even if the solution is nonspinning. Furthermore the solution contains also moments with $m\neq0$, consistently with the fact that 
axisymmetry is broken and reduced to the (discrete) dihedral symmetry $D_3 = Z_3 \rtimes Z_2 = S_3$.

\subsubsection{Solution~${\bf B}$: $\vec\kappa=(1,0,\bar{\kappa}_3,\bar{\kappa}_4\lambda)$}

We consider a non-scaling solution with parameters
\begin{align}
\kappa_i&=\,(1,0,1, \lambda)
\quad, \quad P_0=\,3
\quad, \quad 
Q_I=(\lambda,\lambda,1)
\quad, \quad 
{\cal M} = \frac{2+\lambda}{2} 
\label{kappaBsubcase} \nn\\
\mu_a &=\left(\frac{1{\,+\,}\lambda}{4},\frac{1{\,+\,}\lambda}{4},\frac{1}{2}\right)
\quad, \quad
j_a=\left(\frac{\lambda-1}{4},\frac{1-\lambda}{4},0 \right) \nn\\
\tilde{a}_i&=\left(0, 1, \frac{\lambda-1}{2 \lambda}\right)
\quad, \quad 
\rho = 1
\quad, \quad
\sigma = \frac{1}{(1 +\frac{\lambda -1}{2\lambda} L)^2}
\,.
\end{align}
Notice that $\rho = 1$, which implies $r_{13}=r_{23}$, therefore the vertices form an isosceles triangle. 
Again since $\tilde{a}_1 = 0$ the parameter $L$ is unbounded.  
In the limit of large $\lambda$ with $L =O( \lambda^0)$ one finds
\begin{equation}
\begin{aligned}
{\cal J}&\approx {L \lambda\over 2(L+2)}   \quad, \quad   \chi \approx {2 L\over (L+2) \lambda} \\
\mathfrak{M}_2& \approx 1+\frac{7-4L-L^2}{2\lambda}  \quad ,\quad 
 \mathfrak{S}_3 \approx 1+\frac{6}{\lambda} \,.
 \end{aligned}
\end{equation}
Notice that, for large $\lambda$, $\chi \ll 1$ for any value of $L$, which is consistent with these solutions being 
microstates of a non-spinning BH. More generally, the non-trivial mass and spin multipole moments
in this limit take the form  
\begin{equation}
\mathcal{M}_{2n\,,0}\approx \frac{\lambda}{2} \left(\frac{L}{2}\right)^{2n}  \quad, \quad
\mathcal{S}_{2n+1\,,0} \approx\frac{\lambda}{2}  \left(\frac{L}{2}\right)^{2n+1}\,.
\end{equation}
 that coincide with those of Kerr metric apart from the missing alternating signs. 
This  is not surprising since in the limit of  large $\lambda$ the mass of two centers is much bigger than the third 
one, 
 so the system looks effectively as a 2-center solution.

\subsubsection{Solution~${\bf C}$: $\vec\kappa=(\bar{\kappa}_1,0,\bar{\kappa}_3\, \lambda,\bar{\kappa}_4\, \lambda)$ }

Here we consider the solution for which $\kappa_2 = 0$ and $\kappa_{3,4}\gg \kappa_1$, with arbitrary $\kappa_1$.  Their 
analytic expressions are cumbersome and we 
present them in Appendix~\ref{app:general}. Here we display the formulae for a given choice of the $\bar{\kappa}$'s:
 \begin{align}
\kappa_i=&\,(3,0,\lambda, 2\lambda)\quad, \quad P_0=3
\quad, \quad 
Q_I=(2\lambda^2,6\lambda^2,3\lambda^2)
\quad, \quad 
{\cal M} = \frac{3+11\lambda^2}{4} \nn\\
\mu_a&=\left(\frac{1+2\lambda^2}{4},\frac{1+6\lambda^2}{4},\frac{1+3\lambda^2}{4}\right)
\quad, \quad
j_a=\left(\frac{\lambda}{4},\frac{\lambda}{4},-\frac{\lambda}{2} \right)\nn\\
\tilde{a}_i&=\left( \frac{1}{6 \lambda^2}, 1, - \frac{1}{12 \lambda^2}\right)
\quad, \quad 
\rho = \frac{1}{(1- \frac{L}{6 \lambda^2})^2}
\quad, \quad
\sigma = \frac{1}{(1- \frac{L}{12 \lambda^2})^2}
\,.
\end{align}
The value of $L_{\rm max}$ in this case is
\begin{equation}
L_\textup{max}= 12 \lambda^2 \left(1- \frac{1}{\sqrt{2}}\right)\,.
\end{equation}
The explicit formulae for the multipole moments are not very illuminating, therefore we consider two subcases with large 
$\lambda$: $L\sim O(1)$ and $L \approx \lambda^2 $.  For these choices one finds
\begin{align}
L=&\,1 : \qquad \chi \approx \frac{4\sqrt{3}}{121}\frac{L}{\lambda^3}
\quad, \quad
\mathfrak{M}_{2}\approx 3 \lambda^2
\quad, \quad
\mathfrak{S}_{3} \approx \frac{49}{8} \lambda^2\nn\\
L=&\,\lambda^2: \qquad  \chi \approx \frac{0.05}{\lambda}
\quad, \quad
\mathfrak{M}_{2} \approx 2.76\, \lambda^2
\quad, \quad
\mathfrak{S}_{3} \approx 3.55 \lambda^2\,.
\end{align}
The same scalings with $\lambda$ are found also for the generic solution presented in Sec.~\ref{app:Cgen}.  In particular, notice that in this case $\mathfrak{M}_{2}$ and $\mathfrak{S}_{3}$ are always much bigger than unity, 
which seems a rather general property of this class of solutions~\cite{Bianchi:2020bxa}. 
  
\subsubsection{Solution~${\bf D}$: $\vec\kappa=(\bar{\kappa}_1,\bar{\kappa}_2\, \lambda,\bar{\kappa}_3\, 
\lambda,\bar{\kappa}_4\, \lambda)$ }

A representative example in this class is given by
  \begin{align}
\kappa_i=&\,(3,\lambda,3\lambda, 4\lambda)  \quad P_0=\,3
\quad, \quad 
Q_I=(8\lambda^2,24\lambda^2,66\lambda^2)
\quad, \quad 
{\cal M} = \frac{3+98\lambda^2}{4} \nn \\
\mu_a &=\left(\frac{1{\,+\,}9\lambda^2}{4},\frac{1{\,+\,}41\lambda^2}{4},\frac{1{\,+\,}48\lambda^2}{4}\right)
\quad, \quad
j_a=\left(\lambda{\,+\,}9\lambda^3,\lambda{\,-\,}9\lambda^3,{\,-\,}2 \lambda \right) \nn\\
\tilde{a}_i=\,&\left( \frac{2}{105 \lambda^2}, \frac{8}{35}, \frac{9 \lambda^2-1}{105  \lambda^2}\right)
\quad, \quad 
\rho = \frac{1}{(1- \frac{2 L}{105 \lambda^2})^2}
\quad, \quad
\sigma = \frac{8^2}{35^2} \frac{1}{(1+ \frac{L}{105}\frac{9 \lambda^2-1}{\lambda^2})^2}
\,.
\end{align}
For this choice of $\kappa_i$ the maximum value of $L$ is
\begin{equation}
L_\text{max} = \frac{3 \lambda ^2}{2 (9 \lambda ^2-1 )} \left(\sqrt{5040 \lambda ^2+1289}-43\right)
\end{equation}
for large $\lambda$ we obtain $L_\text{max} \approx 2 \sqrt{35} \lambda\sim 11.8 \lambda$. Again we consider two 
subcases with large $\lambda$:  
\begin{equation}
\begin{aligned}
L=1: \qquad  &\chi \approx \frac{0.003}{\lambda}
\quad, \quad
\mathfrak{M}_2 \approx -\frac{20}{\lambda^2}
\quad, \quad
 \mathfrak{S}_3 \approx -\frac{ 57}{\lambda^2}\\
L=10 \lambda: \qquad  & \chi \approx \frac{0.04}{\lambda}
\quad, \quad
\mathfrak{M}_2\approx 6.9
\quad, \quad
\mathfrak{S}_3  \approx 20
\end{aligned}
\end{equation}

The same scalings with $\lambda$ are found also for the more general solution of this class presented in 
Sec.~\ref{app:k2neq0a}.
In particular, notice that when $\kappa_2\neq0$ the behavior of $\mathfrak{M}_2$ and $\mathfrak{S}_3$ is drastically 
different: in the large-$\lambda$ limit they tend to vanish when $L={\cal O}(1)$, whereas they asymptote to a constant 
value in the opposite regime $L\to L_{\rm max}\sim \lambda$. In all cases, the dimensionless spin $\chi$ is vanishingly 
small.

\subsubsection{Solution~${\bf E}$: $\vec\kappa=(\bar{\kappa}_1 \lambda,\bar{\kappa}_2\, \lambda,\bar{\kappa}_3\, 
\lambda,\bar{\kappa}_4\, \lambda)$ }
 
 A representative example in this class is given by
 \be
\begin{aligned}
\kappa=&\,(2 \lambda,\lambda,4\lambda, 3\lambda) \quad,\quad P_0=\,3
\quad, \quad 
Q_I=(9\lambda^2,18\lambda^3,66\lambda^3)
\quad, \quad 
{\cal M} = \frac{3(1+3\lambda^2+28\lambda^3)}{4} \\
\mu_a & 
=\left(\frac{1{\,+\,}12\lambda^2{\,+\,}2\lambda^3}{4},\frac{1{\,-\,}3\lambda^2{\,+\,}32\lambda^3}{4},\frac{1{\,+\,}
50\lambda^3}{4}\right)
\\
j_a &=\left(\frac{\lambda}{4}   \left(24 \lambda ^3+2 \lambda -1\right),\lambda  \left(-6 \lambda ^3+2 \lambda 
-1\right),\frac{5\lambda}{4} (1-2 \lambda )  \right)
\\
\tilde{a}_i=\,&\left( \frac{5(2\lambda-1)}{324 \lambda^3}, \frac{1}{3}, \frac{6 \lambda^3-2\lambda+1}{81 
\lambda^3}\right)
\quad, \quad 
\rho = \frac{1}{(1- \frac{5(2\lambda-1)}{324 \lambda^3}L)^2}
\quad, \quad
\sigma = \frac{1}{3} \frac{1}{(1+ \frac{6 \lambda^3-2\lambda+1}{81 \lambda^3} L)^2}
\,.
\end{aligned}
\ee
The exact value of $L_{\rm max}$ is not so illuminating therefore we show the large $\lambda$ limit
\begin{equation}
L \leq L_\text{max} \approx \frac{27}{\sqrt{5}} \lambda \sim 12 \lambda
\end{equation}
Again we consider two subcases with large $\lambda$: $L\sim O(1)$ and $L\sim 10\lambda< L_\textup{max}$. One finds
\begin{equation}
\begin{aligned}
L=1:& \qquad \chi \approx {0.004 \over \lambda^2}
\quad, \quad
\mathfrak{M}_{2}  \approx - {14 \over \lambda^2}
\quad, \quad
 \mathfrak{S}_{3} \approx -{61 \over \lambda^2}\\
L=10\lambda: &\qquad \chi \approx {0.06 \over \lambda^2}
\quad, \quad
 \mathfrak{M}_{2}  \approx 2.9
\quad, \quad
 \mathfrak{S}_{3} \approx 13
\end{aligned}
\end{equation}
The same scalings with $\lambda$ are found also for the more general solution of this class presented in 
Sec.~\ref{app:k2neq0b}. Similarly to Solution~{\bf D} above, also in this case
$\mathfrak{M}_2$ and $\mathfrak{S}_3$ vanish in the large-$\lambda$ limit when $L={\cal 
O}(1)$, whereas they asymptote to a constant value in the opposite regime $L\to L_{\rm max}\sim \lambda\gg1$ limit. 
%
%

\subsection{A statistical approach}
\begin{figure}[t]
\centering
 \includegraphics[scale=0.5]{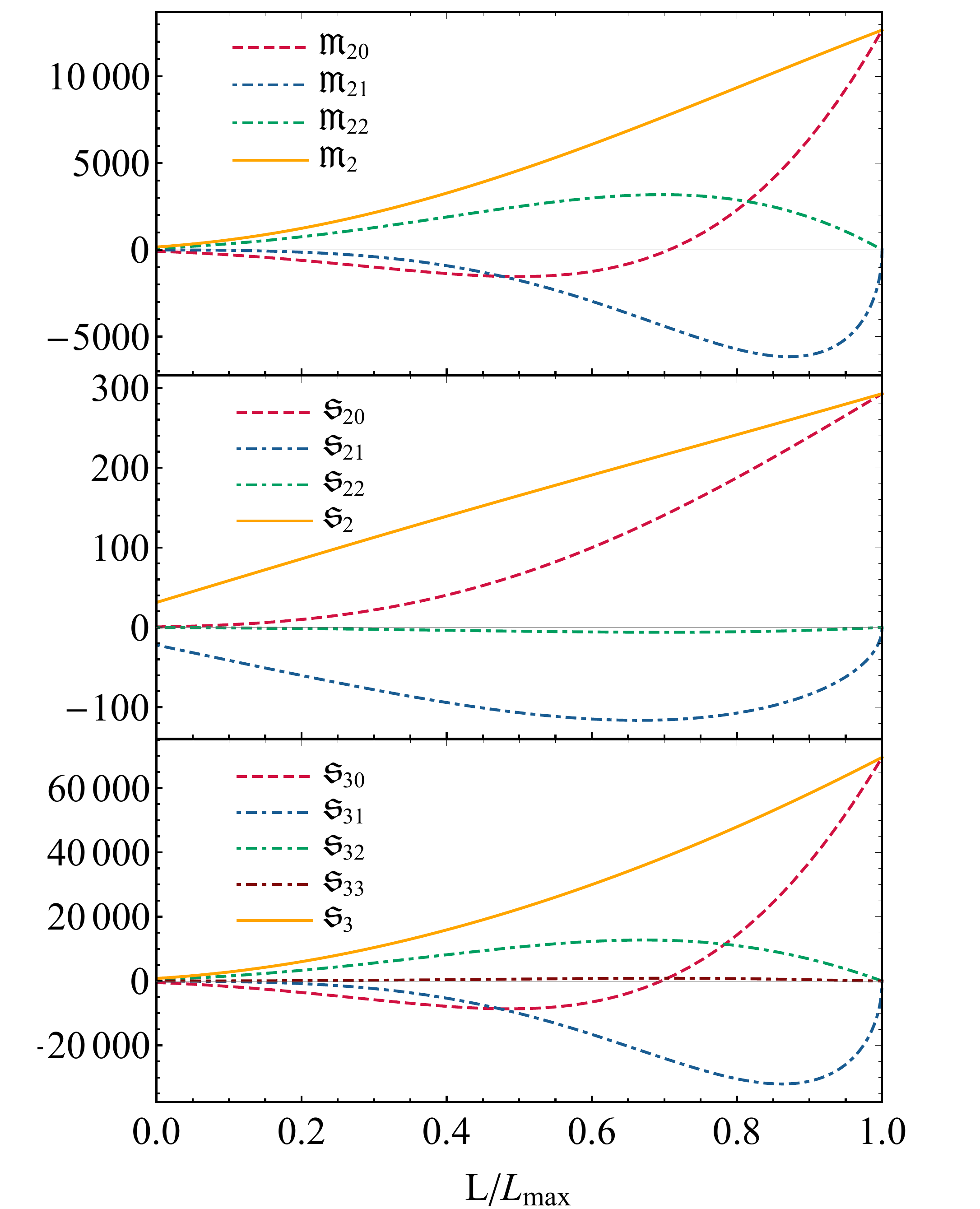} 
\caption{
The quantities $\mathfrak{M}_2$ (top panel), $\mathfrak{S}_2$ (middle panel), and $\mathfrak{S}_3$ (bottom 
panel) defined in Eq.~\eqref{invariants} and the normalized components of the corresponding moments for a 
representative choice of $\kappa_i=(325, 751, 798, 272)$ (corresponding to $L_{\rm max}=79.3361$) as a function 
of $L/ L_{\rm max}\leq1$.
In the top 
and bottom panels the horizontal thin black line refers to the fuzzball and Kerr moments being identical, whereas 
$\mathfrak{S}_{2m}$ are identically zero for Kerr.
All quantities are larger than their Kerr counterpart when $L\sim L_{\rm max}$ while in this example some can be smaller 
when $L\ll L_{\rm max}$. In the $L\to0$ limit the normalized quantities tend to some small but nonvanishing values.
}
\label{fig:examples}
\end{figure}

As clear from the previous sections, even in the simplest family of microstate geometries (with three centers), the 
parameter space is very complex and it is hard to extract general properties from particular classes of solutions.
Nonetheless, our partial exploration of certain classes of solutions suggests the following trend:
\begin{itemize}
 \item In certain subspaces of the parameters (in particular when $\kappa_2=0$), the solutions have 
generically multipole moments larger (in absolute value) than their Kerr counterpart, except for few isolated examples, 
whose measure is of lower dimension relative to the subspace.
 \item In general (\textit{i.e.}, if all $\kappa_i\neq0$) there exists a critical value $L_{\rm crit}$ such that the solutions 
with $L>L_{\rm crit}$ have multipole 
moments larger (in absolute value) than their Kerr counterpart, whereas the opposite is true for $L<L_{\rm crit}$. 
The value of $L_{\rm crit}$ depends on the specific combination of $\kappa_i$ and might also be zero, \textit{i.e.} some 
solutions have larger moments for any $L>0$, as in the previous case.
\end{itemize}
A representative example of these different behaviors is presented in Fig.~\ref{fig:examples}.

To gain some further insight and check these trends, we apply the method presented in the previous sections to compute 
the multipole moments of general solutions found by {\it randomly} selecting the parameters $L$ and $\kappa_i$ (with 
$i=1,2,3,4$). In particular, we draw $10^4$ realizations from a uniform distribution
\begin{equation}
 \kappa_i\in[1,\kappa_{\rm max}]\,,
\end{equation}
constrained by imposing that the conditions in Eq.~\eqref{general_kappa} be satisfied. As a 
representative case we choose $\kappa_{\rm max}=1000$. For a given choice of $\kappa_i$, we draw $L$ from a uniform 
distribution
\begin{equation}
 L\in(0,L_{\rm max}]\,,
\end{equation}
where $L_{\rm max}$ is given in \eqref{Lmax}, or is chosen to be $L_{\rm max}=10^4$ for those 
(few) isolated cases in which $L$ is unbounded.

We find two main results:  

\begin{enumerate}
 \item The normalized invariant $\mathfrak{M}_{2}$ defined in \eqref{invariants} is bigger than its Kerr value 
($\mathfrak{M}_{2}^{\rm Kerr}=1$) for about $90\%$ of the solutions. Similar (slightly higher) percentages apply also 
to higher-order moments and, in particular, to $\mathfrak{S}_{3}$.  These percentages do not depend on the choice of 
$\kappa_{\rm max}$, suggesting that both $L_{\rm crit}$ and $L_{\rm max}$ grow linearly with 
$\kappa_i\gg1$.
 \item For each random realization of $\kappa_i$, the normalized invariants $\mathfrak{M}_{\ell} $ and 
 $\mathfrak{S}_{\ell}$ are {\it always bigger} than their value in the (non-rotating) BH 
limit, \textit{i.e}. when $L\to0$, even when the corresponding moments are not defined in that 
limit~\cite{Bena:2020see,Bena:2020uup}. Indeed, we numerically find that these quantities are always 
monotonous functions of $L$, attaining a global minimum at $L=0$. Note that this property holds only for the 
specific invariants defined by a specific combination of the components of each moment (\textit{e.g.}, 
Eq.~\eqref{invariants}, see also 
Appendix~\ref{app:invariants}) and not for the individual components of the moments (\textit{e.g.}, $\mathfrak{M}_{\ell0}$ and 
$\mathfrak{S}_{\ell0}$ as defined in Eq.~\eqref{mathfrak}).
\end{enumerate}

\section{Conclusions and discussion}
We have extended and refined a general method to determine the multipole moments of spacetimes with a single timelike 
Killing vector field and no extra symmetry. In particular, this technique is useful in the study of the moments of 
fuzzball microstate geometries. These typically break the axial and equatorial symmetries the Kerr metric and are also 
rather complicated. We focused on three-center solutions, but our analysis can be straightforwardly applied to generic 
multi-center solutions and generic BH solutions.

The multipolar structure of fuzzballs is significantly richer than that of a BH in GR, in particular multipole 
moments ${\cal M}_{\ell m}$ and ${\cal S}_{\ell m}$ with $m\neq 0$ (associated with the absence of axial and 
equatorial symmetry) are non-zero, at variance with the Kerr metric~\cite{Bianchi:2020bxa}.

All astrophysical observations so far are perfectly consistent with the hypothesis that all dark compact objects in the 
universe can be described by the Kerr metric~\cite{Cardoso:2019rvt}. Thus, from a phenomenological point of view an 
interesting problem is to understand whether current and future observations can distinguish the classical Kerr metric 
from other 
paradigms, such as the fuzzball proposal. Here we compared the multipole moments of a large family of smooth 
horizonless geometries  with those of a BH with the same mass and spin.

Another natural question would be to compare the multipolar structure of individual microstates with the one of the 
corresponding BH that should emerge from their ensemble. Unfortunately, to the best of our knowledge no fuzzball solutions in four dimensions are known beyond the BPS case, and supersymmetric BHs are necessarily non-spinning in four dimensions~\cite{Townsend:2002yf}. Still, a comparison between the individual microstate and its 
Schwarzschild (or Reissner-Nordstr\"om) BH limit can be performed for some dimensionless ratios which are finite in the 
BH limit. The universal properties of these ratios in the context of BH microstates has been recently studied 
in~\cite{Bena:2020see,Bena:2020uup}.  We find strong numerical evidence that these ratios grow monotonically with the 
microstate 
size $L$, attaining a minimum at the BH limit, $L\to 0$. 
A similar study has been performed in the past for exotic compact objects (\textit{e.g.}, gravastars and 
anisotropic stars)~\cite{Yagi:2015upa,Pani:2015tga,Uchikata:2015yma,Uchikata:2016qku,Raposo:2018xkf} suggesting the 
universal character of this property. 
We also find that the maximum size $L_{\rm max}$ is always 
smaller than the horizon length scale $r_H\sim(Q_1Q_2Q_3 P_0)^{1/4}$, and that $L_{\rm max}\ll r_H$ for large charges. 
In this limit the dimensionless spin $\chi={\cal J}/{\cal M}^2$ is always small, consistently with the fact  the 
solutions represent microstates of a non-rotating BH.

Although the study of the multipole moments of microstate geometries has just 
started \cite{Bena:2020see,Bianchi:2020bxa,Bena:2020uup} and can be extended in various directions, some 
intriguing generic properties seem to appear,  such as the fact that the BH metric seems to be the 
solution with a given mass and spin that typically minimizes the multipole moments or certain combinations thereof.
Indeed, we found that the invariant built from the dimensionless multipole moments of the Kerr metric are smaller than 
those of a given microstate with the same mass and spin in approximately $90\%$ of the four-dimensional parameter space 
of three-center solutions.

It is also intriguing to note that the Lyapunov exponent of 
unstable null geodesics near the photon sphere was found to be maximum for the BH solution relative 
to the microstate geometries~\cite{Bianchi:2020des}. This suggests that the BH metric is an {\it extremum} in the 
parameter space of the solutions of the theory for several (apparently disconnected) quantities.
Clearly, some interesting extensions of our work are to find an analytical proof of the monotonous behaviour of 
$\mathfrak{M}_\ell$ and $\mathfrak{S}_\ell$, and to check whether the above properties hold true also for other 
multi-center microstate geometries.

 We stress the fact that our method can be directly  applied to non-BPS microstate solutions when such solutions would 
be available. In this case microstate  geometries with $\chi={\cal O}(1)$
should exist.
Such an analysis might help understanding how the multipole moments of a classical BH could emerge from an averaging of 
an ensemble of microstates, each microstate having a different multipolar structure.

There is a long way to go before observationally imprints of fuzzballs in astrophysical systems can be modelled 
accurately and, in this respect, several interesting extensions of this and recent 
studies~\cite{Bena:2020see,Bianchi:2020bxa,Bena:2020uup} are urgent. Nonetheless, we believe 
that the analysis of the multipole moments can provide a new portal to constrain fuzzball models with current and future 
observations, by means of both electromagnetic and gravitational-wave probes.

\vspace{1cm}
\noindent{{\bf{\em Note added.}}}
While this work was in preparation, a related work by Iosif Bena and Daniel R.~Mayerson 
appeared~\cite{Bena:2020uup}. Ref.~\cite{Bena:2020uup} and our work are the longer companions to 
Refs.~\cite{Bena:2020see} and \cite{Bianchi:2020bxa}, respectively.

\section*{Acknowledgments}
We thank Iosif Bena and Daniel R.~Mayerson for interesting discussions and for sharing their draft~\cite{Bena:2020uup} 
with us before submission. D.C. was supported by FWF Austrian Science Fund via the SAP P30531-N27.
P.P. acknowledges financial support provided under the European Union's H2020 ERC, Starting 
Grant agreement no.~DarkGRA--757480, and under the MIUR PRIN and FARE programmes (GW-NEXT, CUP:~B84I20000100001), and 
support from the Amaldi Research Center funded by the MIUR program ``Dipartimento di Eccellenza'' (CUP: 
B81I18001170001).
%
%
%
\appendix

\section{Invariants associated to multipole moments} \label{app:invariants}

Using the Cartesian description of the multipoles we can construct some quantities which are invariant under rotations. 
We start from the general formula which connects the Cartesian and the spherical descriptions, 
\begin{equation}
{1\over \ell!} {\cal Q}^{i_1 \dots i_\ell} n_{i_1} \dots n_{i_\ell} = \sqrt{4\pi \over 2\ell+1} \sum_{m=-\ell}^{+\ell} 
\mathcal{M}_{\ell m} Y_{\ell m} 
\label{multi_cartVSsph}
\end{equation}
where ${\cal Q}^{i_1 \dots i_\ell}$ is a symmetric traceless tensor. We now specialize our 
computation to the cases $\ell=1,2,3$ ($\ell=0$ is trivial).

\paragraph{Dipole moments.}

For $\ell=1$, ${\cal Q}_1^{i}$ is a vector. Using \eqref{multi_cartVSsph} we can write the components of ${\cal Q}^{i}$ 
in terms of ${\cal M}_{1m}$:
\begin{equation}
{\cal Q}_1^x =\sqrt{2}({\cal M}_{1-1}-{\cal M}_{11}) \quad, \quad
{\cal Q}_1^y =-i \sqrt{2}({\cal M}_{1-1}+{\cal M}_{11})\quad, \quad
{\cal Q}_1^z =2 {\cal M}_{10}\,.
\end{equation}
The only invariant associated to ${\cal Q}_1^{i}$ is
\begin{equation}
\frac{1}{4} |\vec{Q}_1|^2= \frac{1}{4} Q_1^i Q_{1 , i} = {\cal M}_{10}^2 +2 |{\cal M}_{11}|^2
\end{equation}

\paragraph{Quadrupole moments.}

For $\ell=2$, ${\cal Q}_2^{ij}$ is a symmetric traceless matrix, therefore there are 5 (real) independent components. 
In terms of the ${\cal M}_{2m}$ we have
\begin{equation}
\begin{aligned}
\mathcal{Q}_2^{xx}&= \sqrt{\frac{3}{2}} ({\cal M}_{22}+{\cal M}_{2-2})-{\cal M}_{20}  \quad, \quad
\mathcal{Q}_2^{yy}= -\sqrt{\frac{3}{2}} ({\cal M}_{22}+{\cal M}_{2-2})-{\cal M}_{20} \\
\mathcal{Q}_2^{xy}&= i\sqrt{\frac{3}{2}} ({\cal M}_{22}{-}{\cal M}_{2-2}) \quad, \quad
\mathcal{Q}_2^{xz}= \sqrt{\frac{3}{2}} ({\cal M}_{2-1}{-}{\cal M}_{21}) \quad, \quad
\mathcal{Q}_2^{yz}= -i \sqrt{\frac{3}{2}} ({\cal M}_{2-1}{+}{\cal M}_{21}) \\
\mathcal{Q}_2^{zz}&=-\mathcal{Q}_2^{xx}-\mathcal{Q}_2^{yy} = 2 {\cal M}_{20}
\end{aligned}\,.
\end{equation}
We can associate two invariants to the $\mathcal{Q}$ matrix, namely its trace and determinant. In terms of 
${\cal M}_{2m}$ they read
\begin{equation}
\begin{aligned}
\frac{1}{6} \text{Tr} {\cal Q}^2 &= {\cal M}_{20}^2 + 2 |{\cal M}_{22}|^2+ 2|{\cal M}_{21}|^2 \\
\frac{1}{2} \text{Det} {\cal Q} &= {\cal M}_{20}^3 +3 {\cal M}_{20} ( |{\cal M}_{21}|^2 - 2|{\cal M}_{22}|^2)+3 
\sqrt{6} \,\text{Re}[{\cal M}_{22}^* 
{\cal M}_{21}^2]
\end{aligned}\,.
\end{equation}

\paragraph{Octupole moments.}

In a similar fashion one can compute the trace invariant associated to ${\cal Q}_3$, which is defined as
\begin{equation}
{1\over 3!} {\cal Q}_3^{ijk}n_i n_j n_k= \sqrt{4\pi \over 7} \sum_{m=-3}^{+3} {\cal M}_{3m} Y_{3m} \,.
\end{equation}
The octupole tensor ${\cal Q}_3^{ijk}$ has 7 independent components. The relations with the ${\cal M}_{3m}$ are the 
following
\begin{equation}
\begin{aligned}
\mathcal{Q}_3^{xxx}&= \frac{3}{2} (\sqrt{5} {\cal M}_{3-3}-\sqrt{3} {\cal M}_{3-1}+\sqrt{3} {\cal M}_{31}-\sqrt{5} 
{\cal M}_{3,3})\\
\mathcal{Q}_3^{xxy}&= -\frac{i}{2} (3 \sqrt{5} {\cal M}_{3-3}-\sqrt{3} {\cal M}_{3-1}-\sqrt{3} {\cal M}_{31}+3 \sqrt{5} 
{\cal M}_{3,3})\\
\mathcal{Q}_3^{xxz}&= \frac{1}{2} (\sqrt{30} {\cal M}_{3-2}-6 {\cal M}_{30}+\sqrt{30} {\cal M}_{32})\\
\mathcal{Q}_3^{xyy}&=\frac{1}{2} (-3 \sqrt{5} {\cal M}_{3-3}-\sqrt{3} {\cal M}_{3-1}+\sqrt{3} {\cal M}_{31}+3 \sqrt{5} 
{\cal M}_{3,3})\\
\mathcal{Q}_3^{xyz}&= -i \sqrt{\frac{15}{2}} ({\cal M}_{3-2}-{\cal M}_{32} )\\
\mathcal{Q}_3^{yyy}&= \frac{3 i}{2} (\sqrt{5} {\cal M}_{3-3}+\sqrt{3} {\cal M}_{3-1}+\sqrt{3} {\cal M}_{31}+\sqrt{5} 
{\cal M}_{3,3} )\\
\mathcal{Q}_3^{yyz}&= -\frac{1}{2} (\sqrt{30} {\cal M}_{3-2}+6 {\cal M}_{30}+\sqrt{30} {\cal M}_{32} )\\
\mathcal{Q}_3^{xzz}&= -\mathcal{Q}_3^{xxx}-\mathcal{Q}_3^{xyy} = 2 \sqrt{3} \left({\cal M}_{3-1}-{\cal M}_{31}\right) \\
\mathcal{Q}_3^{yzz}&= -\mathcal{Q}_3^{xxy}-\mathcal{Q}_3^{yyy} = -2 i \sqrt{3} \left({\cal M}_{3-1}+{\cal M}_{31}\right) 
\\
\mathcal{Q}_3^{zzz}&= -\mathcal{Q}_3^{xxz}-\mathcal{Q}_3^{yyz} = 6 {\cal M}_{30}
\end{aligned}\,.
\end{equation}
We can compute $\text{Tr} {\cal Q}_3^2= {\cal Q}_3^{ijk} {\cal Q}_{3, \,ijk}$ obtaining
\begin{equation}
\frac{1}{90} \text{Tr} {\cal Q}_3^2= {\cal M}_{30}^2 + 2 |{\cal M}_{33}|^2+ 2|{\cal M}_{32}|^2+ 2 |{\cal M}_{31}|^2\,.
\end{equation}
 
 \section{Multipole moments of general classes of microstate geometries} \label{app:general}

In this appendix we provide the multipole moments for some generic class of solutions. As discussed in the main text 
the parameter $L$ is generically bounded, $L\leq L_{\rm max}$, except for some particular solutions. Since in 
the large-$\kappa$ limit typically $L_{\rm max}={\cal O}(\kappa)$, we shall distinguish between two opposite regimes:
\begin{enumerate}
 \item $L$ is much smaller than the leading parameter(s) $\kappa_i$ (this includes the small-$L$ limit);
 \item $L\sim L_{\rm max}={\cal O}(\kappa)$.
\end{enumerate}

\subsection{General solution with $\kappa_1={\cal O}(1)$, $\kappa_2=0$, $\kappa_3={\cal O}(1)$,  $\kappa_{4}\gg1$} 
\label{app:Bgen}

We consider the case in which $\kappa_2 = 0$, $ \kappa_{1,3}\sim{\cal O}(1)$ and $\kappa_{4}\gg \kappa_{1,3}$. To 
simplify the notation we 
define $(\kappa_1,\kappa_3,\kappa_4)=(\bar{\kappa}_1,\bar{\kappa}_3,\bar{\kappa}_4 \lambda)$, where 
$\bar{\kappa_i}\sim {\cal O}(1)$ and $\lambda\gg 1$. 

\subsubsection{Small $L$}

If $L\ll\lambda$, to leading order in $\lambda$ the centers are located at
\begin{align}
\textbf{x}_1 &= \left(-\frac{\left(\bar{\kappa} _1^2 \bar{\kappa} _3^2-2 \bar{\kappa} _1 \bar{\kappa} _3^2-1\right) L \sqrt{3 \bar{\kappa} _1^2 \bar{\kappa} _3^4+L^2+4 \bar{\kappa} _1 \bar{\kappa} _3^2 L}}{\left(\bar{\kappa} _1+1\right) \bar{\kappa} _3 \bar{\kappa} _4 \lambda  \left(2 \bar{\kappa} _1 \bar{\kappa} _3^2+L\right)},0,\frac{2 \bar{\kappa} _1^2 \bar{\kappa} _3^2 L}{\left(\bar{\kappa} _1+1\right) \left(2 \bar{\kappa} _1 \bar{\kappa} _3^2+L\right)}\right), \\  
\textbf{x}_2 &= \left(\frac{\left(\left(2 \bar{\kappa} _1-1\right) \bar{\kappa} _3^2+1\right) L \sqrt{3 \bar{\kappa} _1^2 \bar{\kappa} _3^4+L^2+4 \bar{\kappa} _1 \bar{\kappa} _3^2 L}}{\left(\bar{\kappa} _1+1\right) \bar{\kappa} _3 \bar{\kappa} _4 \lambda  \left(2 \bar{\kappa} _1 \bar{\kappa} _3^2+L\right)},0,-\frac{2 \bar{\kappa} _1 \bar{\kappa} _3^2 L}{\left(\bar{\kappa} _1+1\right) \left(2 \bar{\kappa} _1 \bar{\kappa} _3^2+L\right)}\right),\\
\textbf{x}_3 &=\left(-\frac{L \sqrt{3 \bar{\kappa} _1^2 \bar{\kappa} _3^4+L^2+4 \bar{\kappa} _1 \bar{\kappa} _3^2 L}}{2 \bar{\kappa} _1 \bar{\kappa} _3^2+L},0,\frac{\left(\bar{\kappa} _1-1\right) \bar{\kappa} _1 \bar{\kappa} _3^2 L}{\left(\bar{\kappa} _1+1\right) \left(2 \bar{\kappa} _1 \bar{\kappa} _3^2+L\right)}\right),
\end{align}
and the first multipole moments of this solution read
\begin{equation}
\begin{aligned}
\mathcal{M}_{00} &= \frac{1}{4} \lambda  \left(\bar{\kappa }_1+1\right) \bar{\kappa }_3 \bar{\kappa }_4
\quad
,
\quad\mathcal{S}_{10}= \frac{\lambda  L \bar{\kappa }_1 \bar{\kappa }_3^2 \bar{\kappa }_4}{4 \bar{\kappa }_1 
\bar{\kappa 
}_3^2+2 L}\,,\\
\mathcal{M}_{22}&={\cal O}\left(\lambda^0\right)\,,\qquad
\mathcal{M}_{21}={\cal O}\left(\lambda^0\right)\,,\qquad
\mathcal{M}_{20} = \frac{\lambda  L^2 \bar{\kappa }_1^3 \bar{\kappa }_3^5 \bar{\kappa }_4}{\left(\bar{\kappa 
}_1+1\right) \left(2 \bar{\kappa }_1 \bar{\kappa }_3^2+L\right){}^2}\,,\\
\mathcal{S}_{22}&={\cal O}\left(\lambda^0\right)\,,\qquad
\mathcal{S}_{21}= {\cal O}\left(\lambda^0\right)\,,\qquad
\mathcal{S}_{20}=\frac{\lambda  L^2 \left(\bar{\kappa }_1-1\right) \bar{\kappa }_1^2 \bar{\kappa }_3^4 \bar{\kappa 
}_4}{\left(\bar{\kappa }_1+1\right) \left(2 \bar{\kappa }_1 \bar{\kappa }_3^2+L\right){}^2}\,.
\end{aligned} \label{momentsGEN1}
\end{equation}

%

\subsection{General solution with $\kappa_1={\cal O}(1)$, $\kappa_2=0$, and $\kappa_{3,4}\gg1$} \label{app:Cgen}

Here we consider the solution for which $\kappa_2 = 0$ and $\kappa_{3,4}\gg \kappa_1$. We 
define $(\kappa_1,\kappa_3,\kappa_4)=(\bar{\kappa}_1,\bar{\kappa}_3 \lambda,\bar{\kappa}_4 \lambda)$, where 
$\bar{\kappa_i}\sim {\cal O}(1)$ and $\lambda\gg 1$.

\subsubsection{Small $L$}

If $L\ll \lambda$, to leading order in the expansion for large $\lambda$, the coordinates of the centers are 
\begin{align}
\textbf{x}_1 &= \frac{1}{{\cal A}}\left(-\left(\sqrt{3} \bar{\kappa }_1 \bar{\kappa }_3 \left(\bar{\kappa }_1 
\left(\bar{\kappa }_3+\bar{\kappa }_4\right)-2 \bar{\kappa }_4\right)\right),0,-\bar{\kappa }_1 \left(\left(\bar{\kappa 
}_1-2\right) \bar{\kappa 
}_3-2 \bar{\kappa }_4\right) \left(\bar{\kappa }_3-\bar{\kappa }_4\right)\right), \\  
\textbf{x}_2 &= \frac{1}{{\cal A}}\left(-\sqrt{3} \bar{\kappa }_3 \left(\bar{\kappa }_4+\bar{\kappa }_1 
\left(\bar{\kappa }_3-2 
\bar{\kappa }_4\right)\right),0,-2 \bar{\kappa }_1^2 \bar{\kappa }_3^2+\bar{\kappa }_1 \bar{\kappa }_3^2+\bar{\kappa 
}_4 
\left(2 \bar{\kappa }_4-\bar{\kappa }_3\right)\right),\\
\textbf{x}_3 &= \frac{1}{{\cal A}}\left(-\sqrt{3} \bar{\kappa }_4 \left(\bar{\kappa }_4+\bar{\kappa }_1 
\left(\bar{\kappa }_4-2 
\bar{\kappa }_3\right)\right),0,\left(\bar{\kappa }_1-1\right) \left(2 \left(\bar{\kappa }_1+1\right) \bar{\kappa 
}_3-\bar{\kappa }_4\right) \bar{\kappa }_4\right),
\end{align}
where the denominator is ${\cal A}=2 \left(\bar{\kappa 
}_4+\bar{\kappa }_1 \left(\bar{\kappa }_3+\bar{\kappa }_4\right)\right)\sqrt{\left(\bar{\kappa }_1^2-\bar{\kappa 
}_1+1\right) \bar{\kappa }_3^2-\left(\bar{\kappa }_1+1\right) \bar{\kappa }_4 \bar{\kappa }_3+\bar{\kappa }_4^2} $. 
Note that the square root in ${\cal A}$ is 
proportional to the angular momentum, which implies that the zero angular momentum limit is singular. Indeed, this 
general solution does not include Solution {\bf A} in the main text. The latter (as well as all solutions in this 
class for which $\mathbfcal{J}=0$) must be studied separately.

To leading order in $\lambda$, the first multipole 
moments of this solution read 
\begin{equation}
\begin{aligned}
\mathcal{M}_{00} &= \frac{1}{4} \lambda ^2 \bar{\kappa }_3 \left(\bar{\kappa }_4+\bar{\kappa }_1 \left(\bar{\kappa 
}_3+\bar{\kappa }_4\right)\right)
\quad
,
\quad\mathcal{S}_{10}= \frac{1}{4} \lambda  L \sqrt{\left(\bar{\kappa }_1^2-\bar{\kappa }_1+1\right) \bar{\kappa 
}_3^2-\left(\bar{\kappa }_1+1\right) \bar{\kappa }_4 \bar{\kappa }_3+\bar{\kappa }_4^2}\,,\\
\mathcal{M}_{22}&=\frac{\lambda ^2 L^2}{{\cal A}^2}\frac{3}{8} \sqrt{\frac{3}{2}}  \bar{\kappa }_1 \bar{\kappa }_3^2 
\bar{\kappa }_4 \left(\bar{\kappa }_4+\bar{\kappa }_1 \left(\bar{\kappa }_3+\bar{\kappa }_4\right)\right) 
\left(\bar{\kappa }_3 \left(\bar{\kappa }_3+\bar{\kappa }_4\right) \bar{\kappa }_1^2+\left(\bar{\kappa }_3^2-6 
\bar{\kappa }_4 \bar{\kappa }_3+\bar{\kappa }_4^2\right) \bar{\kappa }_1+\bar{\kappa }_4 \left(\bar{\kappa 
}_3+\bar{\kappa }_4\right)\right)\,,\\
\mathcal{M}_{21}&=\frac{\lambda ^2 L^2}{{\cal A}^2}\frac{9}{4 \sqrt{2}}  \left(\bar{\kappa }_1-1\right) \bar{\kappa }_1 
\bar{\kappa }_3^2 \left(\bar{\kappa }_3-\bar{\kappa }_4\right) \left(\bar{\kappa }_1 \bar{\kappa }_3-\bar{\kappa 
}_4\right) \bar{\kappa }_4 \left(\bar{\kappa }_4+\bar{\kappa }_1 \left(\bar{\kappa }_3+\bar{\kappa 
}_4\right)\right)\,,\\
\mathcal{M}_{20} &= \frac{\lambda ^2 L^2}{{\cal A}^2}\frac{1}{8}  \bar{\kappa }_1 \bar{\kappa }_3 \bar{\kappa }_4 
\left(\left(8 \bar{\kappa }_1^3-9 \bar{\kappa }_1^2-9 \bar{\kappa }_1+8\right) \bar{\kappa }_3^3-3 \left(3 \bar{\kappa 
}_1^2-10 \bar{\kappa }_1+3\right) \bar{\kappa }_4 \bar{\kappa }_3^2-9 \left(\bar{\kappa }_1+1\right) \bar{\kappa }_4^2 
\bar{\kappa }_3\right.\\
&\left.+8 \bar{\kappa }_4^3\right) \left(\bar{\kappa }_4+\bar{\kappa }_1 \left(\bar{\kappa }_3+\bar{\kappa 
}_4\right)\right)\,,\\
\mathcal{S}_{22}&=-\frac{\lambda  L^2}{{\cal A}^2} \frac{3}{8}  \sqrt{\frac{3}{2}}  \left(\bar{\kappa }_1-1\right) 
\bar{\kappa }_3 \left(\bar{\kappa }_3-\bar{\kappa }_4\right) \left(\bar{\kappa }_1 \bar{\kappa }_3-\bar{\kappa 
}_4\right) \left(\bar{\kappa }_4+\bar{\kappa }_1 \left(\bar{\kappa }_3+\bar{\kappa }_4\right)\right){}^2\,,\\
\mathcal{S}_{21}&= \frac{\lambda  L^2}{{\cal A}^2}\frac{3}{4\sqrt{2}}  \bar{\kappa }_3 \left(\bar{\kappa 
}_4+\bar{\kappa 
}_1 \left(\bar{\kappa }_3+\bar{\kappa }_4\right)\right) \left(\bar{\kappa }_3 \left(\bar{\kappa }_3^2-4 \bar{\kappa }_4 
\bar{\kappa }_3+\bar{\kappa }_4^2\right) \bar{\kappa }_1^3+\left(\bar{\kappa }_3^3+2 \bar{\kappa }_4 \bar{\kappa 
}_3^2+2 
\bar{\kappa }_4^2 \bar{\kappa }_3+\bar{\kappa }_4^3\right) \bar{\kappa }_1^2\right.\\
& \left. +2 \bar{\kappa }_4 \left(-2 \bar{\kappa }_3^2+\bar{\kappa }_4 \bar{\kappa }_3-2 \bar{\kappa }_4^2\right) 
\bar{\kappa }_1+\bar{\kappa }_4^2 \left(\bar{\kappa }_3+\bar{\kappa }_4\right)\right)\,,\\
\mathcal{S}_{20}&=\frac{\lambda  L^2}{{\cal A}^2} \frac{1}{8} \left(\bar{\kappa }_4+\bar{\kappa }_1 \left(\bar{\kappa 
}_3+\bar{\kappa }_4\right)\right) \left(8 \bar{\kappa }_3^3 \left(\bar{\kappa }_3-\bar{\kappa }_4\right) \bar{\kappa 
}_1^4-7 \left(\bar{\kappa }_3^4-\bar{\kappa }_3^2 \bar{\kappa }_4^2\right) \bar{\kappa }_1^3+7 \left(\bar{\kappa 
}_3^4-\bar{\kappa }_3 \bar{\kappa }_4^3\right) \bar{\kappa }_1^2\right.\\
& \left.-8 \left(\bar{\kappa }_3^4-\bar{\kappa }_4^4\right) \bar{\kappa }_1+\bar{\kappa }_4 \left(8 \bar{\kappa }_3^3-7 
\bar{\kappa }_4 \bar{\kappa }_3^2+7 \bar{\kappa }_4^2 \bar{\kappa }_3-8 \bar{\kappa }_4^3\right)\right)\,,
\end{aligned} \label{momentsGEN2}
\end{equation}
Notice that in the denominator of each of the multipoles there is a term proportional to the angular momentum.

\subsection{General solution with $\kappa_1={\cal O}(1)$ and $\kappa_{2,3,4}\gg 1$} \label{app:k2neq0a}

An even more general solution with $\kappa_2\neq0$ can be constructed analytically when 
$(\kappa_1,\kappa_2,\kappa_3,\kappa_4)=(\bar{\kappa}_1,\bar{\kappa}_2\lambda,\bar{\kappa}_3 \lambda,\bar{\kappa}_4 
\lambda)$, where $\bar{\kappa_i}\sim {\cal O}(1)$ and $\lambda\gg 1$. 

\subsubsection{Small $L$}

If $L\ll \lambda$, to leading order in $\lambda\gg1$ the first multipole moments of this class of solutions are

\begin{equation}
\begin{aligned}
\mathcal{M}_{00} &= \frac{1}{4} \lambda ^2 \left(\bar{\kappa }_1 \bar{\kappa }_2^2+4 \bar{\kappa }_1 \bar{\kappa }_3 
\bar{\kappa }_2-\bar{\kappa }_1 \bar{\kappa }_4 \bar{\kappa }_2-\bar{\kappa }_4 \bar{\kappa }_2+\bar{\kappa }_1 
\bar{\kappa }_3^2+\bar{\kappa }_1 \bar{\kappa }_3 \bar{\kappa }_4+\bar{\kappa }_3 \bar{\kappa }_4\right)\,,\\
\mathcal{S}_{10}&= \frac{\lambda ^3 L \bar{\kappa }_1 \bar{\kappa }_2 \left(\bar{\kappa }_2-\bar{\kappa }_3\right){}^2 
\bar{\kappa }_3 \bar{\kappa }_4}{4 \bar{\kappa }_2^2+4 \bar{\kappa }_3^2+2 (L+5) \bar{\kappa }_3 \bar{\kappa }_2}\,,\\
\mathcal{M}_{22}&=\frac{\lambda ^2 L^2}{\cal Z} \sqrt{\frac{3}{2}} \bar{\kappa }_1 \left(\bar{\kappa }_2+\bar{\kappa 
}_3\right){}^2 \left(\left(\bar{\kappa }_3-\bar{\kappa }_2\right) \bar{\kappa }_4+\bar{\kappa }_1 \left(\bar{\kappa }_2 
\left(2 \bar{\kappa}_3-\bar{\kappa }_4\right)+\bar{\kappa }_3 \bar{\kappa }_4\right)\right) \left(3 \bar{\kappa }_2^4+3 
\bar{\kappa }_3^4\right.\\
& \left.+\left(L^2+10 L+27\right) \bar{\kappa }_3^2 \bar{\kappa }_2^2+4 (L+6) \bar{\kappa }_3 \bar{\kappa }_2^3+4 (L+6) 
\bar{\kappa }_3^3 \bar{\kappa }_2\right)\,,\\
\mathcal{M}_{21}&=-\frac{\lambda^2 L^2}{\cal Z}\sqrt{6}\left(\bar{\kappa }_1-1\right) \bar{\kappa }_1 \left(\bar{\kappa 
}_2-\bar{\kappa }_3\right){}^2 \left(\bar{\kappa }_2+\bar{\kappa }_3\right){}^3 \bar{\kappa }_4\left(3 \bar{\kappa 
}_2^4+3 \bar{\kappa }_3^4+\left(L^2+10 L+27\right) \bar{\kappa }_3^2 \bar{\kappa }_2^2\right.\\
&\left.+4 (L+6) \bar{\kappa }_3 \bar{\kappa }_2^3+4 (L+6) \bar{\kappa }_3^3 \bar{\kappa }_2\right)^{1/2} \,, \\
\mathcal{M}_{20} &= -\frac{\lambda^2 \-L^2}{{\cal Z}}\left[8 \bar{\kappa }_2 \bar{\kappa }_3 \bar{\kappa }_4^2 
\left(\bar{\kappa }_2-\bar{\kappa }_3\right){}^4-\bar{\kappa }_1 \bar{\kappa }_4 \left(\bar{\kappa }_2-\bar{\kappa 
}_3\right) \left(\bar{\kappa }_2^6+\bar{\kappa }_3^5 \left(\bar{\kappa }_3-8 \bar{\kappa }_4\right)\right.\right.\\
&+\bar{\kappa }_3 \bar{\kappa }_2^4 \left(\left(L^2+18 L+112\right) \bar{\kappa }_3-24 \bar{\kappa }_4\right)+2 
\bar{\kappa }_3^2 \bar{\kappa }_2^3 \left(16 \bar{\kappa }_4+\left(L^2+14 L+23\right) \bar{\kappa }_3\right)\\
 & \left.+\bar{\kappa }_3^3 \bar{\kappa }_2^2 \left(\left(L^2+18 L+112\right) \bar{\kappa }_3-32 \bar{\kappa 
}_4\right)+\bar{\kappa }_2^5 \left(8 \bar{\kappa }_4+(4 L+26) \bar{\kappa }_3\right)+2 \bar{\kappa }_3^4 \bar{\kappa 
}_2 
\left(12 \bar{\kappa }_4+(2 L+13) \bar{\kappa }_3\right)\right)\\
& +\bar{\kappa }_1^2 \left(\left(2 \bar{\kappa }_3-\bar{\kappa }_4\right) \bar{\kappa }_2^7+\bar{\kappa }_3^7 
\bar{\kappa }_4+\bar{\kappa }_3 \bar{\kappa }_2^5 \left(8 \bar{\kappa }_4^2+2 \left(L^2+18 L+96\right) \bar{\kappa 
}_3^2-\left(L^2+14 L+86\right) \bar{\kappa }_4 \bar{\kappa }_3\right)\right.\\
 & +\bar{\kappa }_3^2 \bar{\kappa }_2^4 \left(-32 \bar{\kappa }_4^2+4 \left(L^2+14 L+35\right) \bar{\kappa 
}_3^2-\left(L^2+10 L-66\right) \bar{\kappa }_4 \bar{\kappa }_3\right)+\bar{\kappa }_3^3 \bar{\kappa }_2^3 \left(48 
\bar{\kappa }_4^2\right.\\
 &  \left. +2 \left(L^2+18 L+96\right) \bar{\kappa }_3^2+\left(L^2+10 L-66\right) \bar{\kappa }_4 \bar{\kappa 
}_3\right)+\bar{\kappa }_3^4 \bar{\kappa }_2^2 \left(-32 \bar{\kappa }_4^2+\left(L^2+14 L+86\right) \bar{\kappa }_4 
\bar{\kappa }_3 \right. \\
 & \left.\left.\left. +(8 L+60) \bar{\kappa }_3^2\right)+\bar{\kappa }_3 \bar{\kappa }_2^6 \left((8 L+60) \bar{\kappa 
}_3-(4 L+25) \bar{\kappa }_4\right)+\bar{\kappa }_3^5 \bar{\kappa }_2 \left(2 \bar{\kappa }_3^2+8 \bar{\kappa }_4^2+(4 
L+25) \bar{\kappa }_4 \bar{\kappa }_3\right)\right)\right]\,,\\
\mathcal{S}_{22}&=0\,,\\
\mathcal{S}_{21}&= \frac{\lambda ^3 L^2 }{\cal Z} 2\sqrt{6} \bar{\kappa }_1^2 \bar{\kappa }_2 \left(\bar{\kappa 
}_2-\bar{\kappa }_3\right){}^2 \bar{\kappa }_3 \left(\bar{\kappa }_2+\bar{\kappa }_3\right){}^2 \bar{\kappa }_4 \left(3 
\bar{\kappa }_2^4+3 \bar{\kappa }_3^4+\left(L^2+10 L+27\right) \bar{\kappa }_3^2 \bar{\kappa }_2^2\right.\\
& \left.+4 (L+6) \bar{\kappa }_3 \bar{\kappa }_2^3+4 (L+6) \bar{\kappa }_3^3 \bar{\kappa }_2\right)^{1/2}\,,\\
\mathcal{S}_{20}&=\frac{ \lambda^3 L^2}{\cal Z} 8 \left(\bar{\kappa }_1-1\right) \bar{\kappa }_1 \bar{\kappa }_2 
\left(\bar{\kappa }_2-\bar{\kappa }_3\right){}^4 \bar{\kappa }_3 \left(\bar{\kappa }_2+\bar{\kappa }_3\right) 
\bar{\kappa }_4^2\,.
\end{aligned}
\end{equation}
where for readability we have defined
\begin{equation}
{\cal Z}=8 \left(\left(\bar{\kappa }_3-\bar{\kappa }_2\right) \bar{\kappa }_4+\bar{\kappa }_1 \left(\bar{\kappa 
}_2^2+\left(4 \bar{\kappa }_3-\bar{\kappa }_4\right) \bar{\kappa }_2+\bar{\kappa }_3 \left(\bar{\kappa }_3+\bar{\kappa 
}_4\right)\right)\right) \left(2 \bar{\kappa }_2^2+2 \bar{\kappa }_3^2+(L+5) \bar{\kappa }_3 \bar{\kappa 
}_2\right){}^2\,.
\end{equation}

\subsubsection{$L\sim L_{\rm max}$}
Since $L\sim L_{\rm max}\sim \lambda$, we can define $L=\bar{L}\lambda$ and, to leading order in $\lambda$, the multipole 
moments read

\begin{equation}
\begin{aligned}
\mathcal{M}_{00} &= \frac{1}{4} \lambda ^2 \left(\left(\bar{\kappa }_3-\bar{\kappa }_2\right) \bar{\kappa 
}_4+\bar{\kappa }_1 \left(\bar{\kappa }_2^2+\left(4 \bar{\kappa }_3-\bar{\kappa }_4\right) \bar{\kappa }_2+\bar{\kappa 
}_3 \left(\bar{\kappa }_3+\bar{\kappa }_4\right)\right)\right)\,,\\
\mathcal{S}_{10}&= \frac{1}{4} \lambda ^2 \bar{L} \left(\bar{\kappa }_1-1\right) \left(\bar{\kappa }_2+\bar{\kappa 
}_3\right)\,,\\
\mathcal{M}_{22}&=0\,,\\
\mathcal{M}_{21}&=0 \,, \\
\mathcal{M}_{20} &= \frac{\lambda ^4 \bar{L}^2 \bar{\kappa }_1 \left(\bar{\kappa }_2+\bar{\kappa }_3\right){}^2 
\left(\left(\bar{\kappa }_3-\bar{\kappa }_2\right) \bar{\kappa }_4+\bar{\kappa }_1 \left(\bar{\kappa }_2 \left(2 
\bar{\kappa }_3-\bar{\kappa }_4\right)+\bar{\kappa }_3 \bar{\kappa }_4\right)\right)}{4 \left(\left(\bar{\kappa 
}_3-\bar{\kappa }_2\right) \bar{\kappa }_4+\bar{\kappa }_1 \left(\bar{\kappa }_2^2+\left(4 \bar{\kappa }_3-\bar{\kappa 
}_4\right) \bar{\kappa }_2+\bar{\kappa }_3 \left(\bar{\kappa }_3+\bar{\kappa }_4\right)\right)\right)}\,,\\
\mathcal{S}_{22}&=0\,,\\
\mathcal{S}_{21}&= 0\,,\\
\mathcal{S}_{20}&= \frac{\lambda ^3 \bar{L}^2 \left(\bar{\kappa }_1-1\right) \left(\bar{\kappa }_2+\bar{\kappa 
}_3\right) \left(\left(\bar{\kappa }_2-\bar{\kappa }_3\right) \bar{\kappa }_4+\bar{\kappa }_1 \left(\bar{\kappa 
}_2^2+\bar{\kappa }_4 \bar{\kappa }_2+\bar{\kappa }_3 \left(\bar{\kappa }_3-\bar{\kappa }_4\right)\right)\right)}{4 
\left(\left(\bar{\kappa }_3-\bar{\kappa }_2\right) \bar{\kappa }_4+\bar{\kappa }_1 \left(\bar{\kappa }_2^2+\left(4 
\bar{\kappa }_3-\bar{\kappa }_4\right) \bar{\kappa }_2+\bar{\kappa }_3 \left(\bar{\kappa }_3+\bar{\kappa 
}_4\right)\right)\right)}\,.
\end{aligned}
\end{equation}
Note that in this case the moments with $m\neq0$ vanish, consistently with the fact that when $L\to L_{\rm max}$ the 
solution is axisymmetric.

\subsection{General solution with $\kappa_{1,2,3,4}\gg1$} \label{app:k2neq0b}
Finally, let us consider the case in which all $\kappa$'s are large, i.e. $\kappa_i=\bar\kappa_i \lambda$ 
($i=1,2,3,4$), with $\bar\kappa_i={\cal O}(1)$ and $\lambda\gg1$.

\subsubsection{Small $L$}

If $L\ll \lambda$, to leading order in $\lambda\gg1$, the multipole moments in this case read

\begin{equation}
\begin{aligned}
\mathcal{M}_{00} &= \frac{1}{4} \lambda ^3 \bar{\kappa }_1 \left(\bar{\kappa }_2^2+\left(4 \bar{\kappa }_3-\bar{\kappa 
}_4\right) \bar{\kappa }_2+\bar{\kappa }_3 \left(\bar{\kappa }_3+\bar{\kappa }_4\right)\right)\,,\\
\mathcal{S}_{10}&= \frac{\lambda ^4 L \bar{\kappa }_1 \bar{\kappa }_2 \left(\bar{\kappa }_2-\bar{\kappa }_3\right){}^2 
\bar{\kappa }_3 \bar{\kappa }_4}{4 \bar{\kappa }_2^2+4 \bar{\kappa }_3^2+2 (L+5) \bar{\kappa }_3 \bar{\kappa }_2}\,,\\
\mathcal{M}_{22}&=\frac{\lambda ^3 L^2}\Upsilon\sqrt{\frac{3}{2}}  \bar{\kappa }_1 \left(\bar{\kappa }_2+\bar{\kappa 
}_3\right){}^2 \left(\bar{\kappa }_2 \left(2 \bar{\kappa }_3-\bar{\kappa }_4\right)+\bar{\kappa }_3 \bar{\kappa 
}_4\right) \left(3 \bar{\kappa }_2^4+3 \bar{\kappa }_3^4+\left(L^2+10 L+27\right) \bar{\kappa }_3^2 \bar{\kappa 
}_2^2\right.\\
&\left.+4 (L+6) \bar{\kappa }_3 \bar{\kappa }_2^3+4 (L+6) \bar{\kappa }_3^3 \bar{\kappa }_2\right)\,,\\
\mathcal{M}_{21}&=-\frac{\lambda ^3 L^2}{\cal Z }\sqrt{6}  \bar{\kappa }_1 \left(\bar{\kappa }_2-\bar{\kappa 
}_3\right){}^2 \left(\bar{\kappa }_2+\bar{\kappa }_3\right){}^3 \bar{\kappa }_4 \left(3 \bar{\kappa }_2^4+3 \bar{\kappa 
}_3^4+\left(L^2+10 L+27\right) \bar{\kappa }_3^2 \bar{\kappa }_2^2\right.\\
&\left.+4 (L+6) \bar{\kappa }_3 \bar{\kappa }_2^3+4 (L+6) \bar{\kappa }_3^3 \bar{\kappa }_2\right)^{1/2} \,, \\
\mathcal{M}_{20} &= -\frac{\lambda ^3 L^2}\Upsilon \bar{\kappa }_1 \left(\left(2 \bar{\kappa }_3-\bar{\kappa 
}_4\right) \bar{\kappa }_2^7+\bar{\kappa }_3^7 \bar{\kappa }_4+\bar{\kappa }_3 \bar{\kappa }_2^5 \left(8 \bar{\kappa 
}_4^2+2 \left(L^2+18 L+96\right) \bar{\kappa }_3^2\right.\right.\\
& \left.-\left(L^2+14 L+86\right) \bar{\kappa }_4 \bar{\kappa }_3\right)+\bar{\kappa }_3^2 \bar{\kappa }_2^4 \left(-32 
\bar{\kappa }_4^2+4 \left(L^2+14 L+35\right) \bar{\kappa }_3^2-\left(L^2+10 L-66\right) \bar{\kappa }_4 \bar{\kappa 
}_3\right)\\
&+\bar{\kappa }_3^3 \bar{\kappa }_2^3 \left(48 \bar{\kappa }_4^2+2 \left(L^2+18 L+96\right) \bar{\kappa 
}_3^2+\left(L^2+10 L-66\right) \bar{\kappa }_4 \bar{\kappa }_3\right)\\
&+\bar{\kappa }_3^4 \bar{\kappa }_2^2 \left(-32 \bar{\kappa }_4^2+\left(L^2+14 L+86\right) \bar{\kappa }_4 \bar{\kappa 
}_3+(8 L+60) \bar{\kappa }_3^2\right)+\bar{\kappa }_3 \bar{\kappa }_2^6 \left((8 L+60) \bar{\kappa }_3-(4 L+25) 
\bar{\kappa }_4\right)\\
&\left.+\bar{\kappa }_3^5 \bar{\kappa }_2 \left(2 \bar{\kappa }_3^2+8 \bar{\kappa }_4^2+(4 L+25) \bar{\kappa }_4 
\bar{\kappa }_3\right)\right)\,,\\
\mathcal{S}_{22}&=0\,,\\
\mathcal{S}_{21}&= \frac{\lambda ^4 L^2}\Upsilon 2 \sqrt{6} \bar{\kappa }_1 \bar{\kappa }_2 \bar{\kappa }_3 
\left(\bar{\kappa }_2^2-\bar{\kappa }_3^2\right){}^2 \bar{\kappa }_4 \left(3 \bar{\kappa }_2^4+3 \bar{\kappa 
}_3^4+\left(L^2+10 L+27\right) \bar{\kappa }_3^2 \bar{\kappa }_2^2\right.\\
&\left.+4 (L+6) \bar{\kappa }_3 \bar{\kappa }_2^3+4 (L+6) \bar{\kappa }_3^3 \bar{\kappa }_2\right)^{1/2}\,,\\
\mathcal{S}_{20}&= \frac{\lambda ^4 L^2}\Upsilon 8\bar{\kappa }_1 \bar{\kappa }_2 \left(\bar{\kappa }_2-\bar{\kappa 
}_3\right){}^4 \bar{\kappa }_3 \left(\bar{\kappa }_2+\bar{\kappa }_3\right) \bar{\kappa }_4^2\,.
\end{aligned}
\end{equation}
where we have defined
\begin{equation}
\Upsilon=8 \left(\bar{\kappa }_2^2+\left(4 \bar{\kappa }_3-\bar{\kappa }_4\right) \bar{\kappa }_2+\bar{\kappa }_3 
\left(\bar{\kappa }_3+\bar{\kappa }_4\right)\right) \left(2 \bar{\kappa }_2^2+2 \bar{\kappa }_3^2+(L+5) \bar{\kappa }_3 
\bar{\kappa }_2\right){}^2\,.
\end{equation}

Notice that this solution can be also obtained from the one in Sec.~\ref{app:k2neq0a} in the $\bar\kappa_1\gg1$ limit.

\subsubsection{$L\sim L_{\rm max}$}

Since $L= \bar{L}\lambda \sim L_{\rm max} $, to leading order in $\lambda$ and with $\bar{L}={\cal O}(1)$, 
the multipole moments read

\begin{equation}
\begin{aligned}
\mathcal{M}_{00} &= \frac{1}{4} \lambda ^3 \bar{\kappa }_1 \left(\bar{\kappa }_2^2+\left(4 \bar{\kappa }_3-\bar{\kappa 
}_4\right) \bar{\kappa }_2+\bar{\kappa }_3 \left(\bar{\kappa }_3+\bar{\kappa }_4\right)\right)\,,\\
\mathcal{S}_{10}&= \frac{1}{4} \lambda ^3 \bar{L} \bar{\kappa }_1 \left(\bar{\kappa }_2+\bar{\kappa }_3\right)\,,\\
\mathcal{M}_{22}&=0\,,\\
\mathcal{M}_{21}&=0 \,, \\
\mathcal{M}_{20} &= \frac{\lambda ^5 \bar{L}^2 \bar{\kappa }_1 \left(\bar{\kappa }_2+\bar{\kappa }_3\right){}^2 
\left(\bar{\kappa }_2 \left(2 \bar{\kappa }_3-\bar{\kappa }_4\right)+\bar{\kappa }_3 \bar{\kappa }_4\right)}{4 
\left(\bar{\kappa }_2^2+\left(4 \bar{\kappa }_3-\bar{\kappa }_4\right) \bar{\kappa }_2+\bar{\kappa }_3 \left(\bar{\kappa 
}_3+\bar{\kappa }_4\right)\right)}\,,\\
\mathcal{S}_{22}&=0\,,\\
\mathcal{S}_{21}&= 0\,,\\
\mathcal{S}_{20}&= \frac{\lambda ^4 \bar{L}^2 \bar{\kappa }_1 \left(\bar{\kappa }_2+\bar{\kappa }_3\right) 
\left(\bar{\kappa }_2^2+\bar{\kappa }_4 \bar{\kappa }_2+\bar{\kappa }_3 \left(\bar{\kappa }_3-\bar{\kappa 
}_4\right)\right)}{4 \left(\bar{\kappa }_2^2+\left(4 \bar{\kappa }_3-\bar{\kappa }_4\right) \bar{\kappa }_2+\bar{\kappa 
}_3 \left(\bar{\kappa }_3+\bar{\kappa }_4\right)\right)}\,,
\end{aligned}
\end{equation}
and also in this case the solution is axisymmetric, as expected.

\bibliography{Ref}

\providecommand{\href}[2]{#2}\begingroup\raggedright\begin{thebibliography}{10}

\bibitem{Carter71}
B.~Carter, \emph{Axisymmetric black hole has only two degrees of freedom},
  \href{http://dx.doi.org/10.1103/PhysRevLett.26.331}{\emph{Phys. Rev. Lett.}
  {\bf 26} (Feb, 1971) 331--333}.

\bibitem{Hawking:1973uf}
S.~Hawking and G.~Ellis, \emph{{The Large Scale Structure of Space-Time}}.
\newblock Cambridge Monographs on Mathematical Physics. Cambridge University
  Press, 2, 2011,
  \href{http://dx.doi.org/10.1017/CBO9780511524646}{10.1017/CBO9780511524646}.

\bibitem{Heusler:1998ua}
M.~Heusler, \emph{{Stationary black holes: Uniqueness and beyond}},
  {\emph{Living Rev. Relativity} {\bf 1} (1998) }.

\bibitem{Chrusciel:2012jk}
P.~T. Chrusciel, J.~L. Costa and M.~Heusler, \emph{{Stationary Black Holes:
  Uniqueness and Beyond}}, {\emph{Living Rev.Rel.} {\bf 15} (2012) 7},
  [\href{http://arxiv.org/abs/1205.6112}{{\tt 1205.6112}}].

\bibitem{Robinson}
D.~Robinson, \emph{{Four decades of black holes uniqueness theorems}}.
\newblock Cambridge University Press, 2009.

\bibitem{Kerr:1963ud}
R.~P. Kerr, \emph{{Gravitational field of a spinning mass as an example of
  algebraically special metrics}},
  \href{http://dx.doi.org/10.1103/PhysRevLett.11.237}{\emph{Phys. Rev. Lett.}
  {\bf 11} (1963) 237--238}.

\bibitem{Geroch:1970cd}
R.~P. Geroch, \emph{{Multipole moments. II. Curved space}},
  \href{http://dx.doi.org/10.1063/1.1665427}{\emph{J.Math.Phys.} {\bf 11}
  (1970) 2580--2588}.

\bibitem{Hansen:1974zz}
R.~Hansen, \emph{{Multipole moments of stationary space-times}},
  \href{http://dx.doi.org/10.1063/1.1666501}{\emph{J.Math.Phys.} {\bf 15}
  (1974) 46--52}.

\bibitem{Pani:2015tga}
P.~Pani, \emph{{I-Love-Q relations for gravastars and the approach to the
  black-hole limit}},
  \href{http://dx.doi.org/10.1103/PhysRevD.95.049902}{\emph{Phys. Rev. D} {\bf
  92} (2015) 124030}, [\href{http://arxiv.org/abs/1506.06050}{{\tt
  1506.06050}}].

\bibitem{Uchikata:2015yma}
N.~Uchikata and S.~Yoshida, \emph{{Slowly rotating thin shell gravastars}},
  \href{http://dx.doi.org/10.1088/0264-9381/33/2/025005}{\emph{Class. Quant.
  Grav.} {\bf 33} (2016) 025005}, [\href{http://arxiv.org/abs/1506.06485}{{\tt
  1506.06485}}].

\bibitem{Uchikata:2016qku}
N.~Uchikata, S.~Yoshida and P.~Pani, \emph{{Tidal deformability and I-Love-Q
  relations for gravastars with polytropic thin shells}},
  \href{http://dx.doi.org/10.1103/PhysRevD.94.064015}{\emph{Phys. Rev. D} {\bf
  94} (2016) 064015}, [\href{http://arxiv.org/abs/1607.03593}{{\tt
  1607.03593}}].

\bibitem{Raposo:2018xkf}
G.~Raposo, P.~Pani and R.~Emparan, \emph{{Exotic compact objects with soft
  hair}}, \href{http://dx.doi.org/10.1103/PhysRevD.99.104050}{\emph{Phys. Rev.
  D} {\bf 99} (2019) 104050}, [\href{http://arxiv.org/abs/1812.07615}{{\tt
  1812.07615}}].

\bibitem{Psaltis:2008bb}
D.~Psaltis, \emph{{Probes and Tests of Strong-Field Gravity with Observations
  in the Electromagnetic Spectrum}},
  \href{http://arxiv.org/abs/0806.1531}{{\tt 0806.1531}}.

\bibitem{Yunes:2013dva}
N.~Yunes and X.~Siemens, \emph{{Gravitational-Wave Tests of General Relativity
  with Ground-Based Detectors and Pulsar Timing-Arrays}},
  \href{http://dx.doi.org/10.12942/lrr-2013-9}{\emph{Living Rev.Rel.} {\bf 16}
  (2013) 9}, [\href{http://arxiv.org/abs/1304.3473}{{\tt 1304.3473}}].

\bibitem{Berti:2015itd}
E.~Berti et~al., \emph{{Testing General Relativity with Present and Future
  Astrophysical Observations}},
  \href{http://dx.doi.org/10.1088/0264-9381/32/24/243001}{\emph{Class. Quant.
  Grav.} {\bf 32} (2015) 243001}, [\href{http://arxiv.org/abs/1501.07274}{{\tt
  1501.07274}}].

\bibitem{Gair:2012nm}
J.~R. Gair, M.~Vallisneri, S.~L. Larson and J.~G. Baker, \emph{{Testing General
  Relativity with Low-Frequency, Space-Based Gravitational-Wave Detectors}},
  \href{http://dx.doi.org/10.12942/lrr-2013-7}{\emph{Living Rev.Rel.} {\bf 16}
  (2013) 7}, [\href{http://arxiv.org/abs/1212.5575}{{\tt 1212.5575}}].

\bibitem{Cardoso:2016ryw}
V.~Cardoso and L.~Gualtieri, \emph{{Testing the black hole no-hair
  hypothesis}},
  \href{http://dx.doi.org/10.1088/0264-9381/33/17/174001}{\emph{Class. Quant.
  Grav.} {\bf 33} (2016) 174001}, [\href{http://arxiv.org/abs/1607.03133}{{\tt
  1607.03133}}].

\bibitem{Barack:2018yly}
L.~Barack et~al., \emph{{Black holes, gravitational waves and fundamental
  physics: a roadmap}},
  \href{http://dx.doi.org/10.1088/1361-6382/ab0587}{\emph{Class. Quant. Grav.}
  {\bf 36} (2019) 143001}, [\href{http://arxiv.org/abs/1806.05195}{{\tt
  1806.05195}}].

\bibitem{Cardoso:2019rvt}
V.~Cardoso and P.~Pani, \emph{{Testing the nature of dark compact objects: a
  status report}},
  \href{http://dx.doi.org/10.1007/s41114-019-0020-4}{\emph{Living Rev. Rel.}
  {\bf 22} (2019) 4}, [\href{http://arxiv.org/abs/1904.05363}{{\tt
  1904.05363}}].

\bibitem{Hertog:2017vod}
T.~Hertog and J.~Hartle, \emph{{Observational Implications of Fuzzball
  Formation}}, \href{http://dx.doi.org/10.1007/s10714-020-02720-z}{\emph{Gen.
  Rel. Grav.} {\bf 52} (2020) 67}, [\href{http://arxiv.org/abs/1704.02123}{{\tt
  1704.02123}}].

\bibitem{Guo:2017jmi}
B.~Guo, S.~Hampton and S.~D. Mathur, \emph{{Can we observe fuzzballs or
  firewalls?}}, \href{http://dx.doi.org/10.1007/JHEP07(2018)162}{\emph{JHEP}
  {\bf 07} (2018) 162}, [\href{http://arxiv.org/abs/1711.01617}{{\tt
  1711.01617}}].

\bibitem{Abbott:2020khf}
{\scshape LIGO Scientific, Virgo} collaboration, R.~Abbott et~al.,
  \emph{{GW190814: Gravitational Waves from the Coalescence of a 23 Solar Mass
  Black Hole with a 2.6 Solar Mass Compact Object}},
  \href{http://dx.doi.org/10.3847/2041-8213/ab960f}{\emph{Astrophys. J.} {\bf
  896} (2020) L44}, [\href{http://arxiv.org/abs/2006.12611}{{\tt 2006.12611}}].

\bibitem{Abbott:2020tfl}
{\scshape LIGO Scientific, Virgo} collaboration, R.~Abbott et~al.,
  \emph{{GW190521: A Binary Black Hole Merger with a Total Mass of $150 ~
  M_{\odot}$}},
  \href{http://dx.doi.org/10.1103/PhysRevLett.125.101102}{\emph{Phys. Rev.
  Lett.} {\bf 125} (2020) 101102}, [\href{http://arxiv.org/abs/2009.01075}{{\tt
  2009.01075}}].

\bibitem{Abbott:2020mjq}
{\scshape LIGO Scientific, Virgo} collaboration, R.~Abbott et~al.,
  \emph{{Properties and astrophysical implications of the 150 Msun binary black
  hole merger GW190521}},
  \href{http://dx.doi.org/10.3847/2041-8213/aba493}{\emph{Astrophys. J. Lett.}
  {\bf 900} (2020) L13}, [\href{http://arxiv.org/abs/2009.01190}{{\tt
  2009.01190}}].

\bibitem{Penrose:1969pc}
R.~Penrose, \emph{{Gravitational collapse: The role of general relativity}},
  {\emph{Riv. Nuovo Cim.} {\bf 1} (1969) 252--276}.

\bibitem{Wald:1997wa}
R.~M. Wald, \emph{{Gravitational collapse and cosmic censorship}},  in
  \emph{{Black Holes, Gravitational Radiation and the Universe: Essays in Honor
  of C.V. Vishveshwara}}, pp.~69--85, 1997.
\newblock \href{http://arxiv.org/abs/gr-qc/9710068}{{\tt gr-qc/9710068}}.
\newblock \href{http://dx.doi.org/10.1007/978-94-017-0934-7_5}{DOI}.

\bibitem{Penrose_CCC}
R.~Penrose, \emph{{Singularities of Spacetime (in Theoretical Principles in
  Astrophysics and Relativity)}},  in \emph{{Chicago University Press, Chicago,
  1978 217 P.}}, 1978.

\bibitem{Bekenstein}
J.~D. Bekenstein, \emph{{Black holes and entropy}}, {\emph{Physical Review D}
  {\bf 7} (1973) 2333}.

\bibitem{Hawking:1976de}
S.~W. Hawking, \emph{{Black Holes and Thermodynamics}},
  \href{http://dx.doi.org/10.1103/PhysRevD.13.191}{\emph{Phys. Rev.} {\bf D13}
  (1976) 191--197}.

\bibitem{Hawking:1974sw}
S.~Hawking, \emph{{Particle Creation by Black Holes}},
  \href{http://dx.doi.org/10.1007/BF02345020}{\emph{Commun. Math. Phys.} {\bf
  43} (1975) 199--220}.

\bibitem{Strominger:1996sh}
A.~Strominger and C.~Vafa, \emph{{Microscopic origin of the Bekenstein-Hawking
  entropy}}, \href{http://dx.doi.org/10.1016/0370-2693(96)00345-0}{\emph{Phys.
  Lett. B} {\bf 379} (1996) 99--104},
  [\href{http://arxiv.org/abs/hep-th/9601029}{{\tt hep-th/9601029}}].

\bibitem{Horowitz:1996ay}
G.~T. Horowitz, J.~M. Maldacena and A.~Strominger, \emph{{Nonextremal black
  hole microstates and U duality}},
  \href{http://dx.doi.org/10.1016/0370-2693(96)00738-1}{\emph{Phys. Lett. B}
  {\bf 383} (1996) 151--159}, [\href{http://arxiv.org/abs/hep-th/9603109}{{\tt
  hep-th/9603109}}].

\bibitem{Maldacena:1997de}
J.~M. Maldacena, A.~Strominger and E.~Witten, \emph{{Black hole entropy in M
  theory}}, \href{http://dx.doi.org/10.1088/1126-6708/1997/12/002}{\emph{JHEP}
  {\bf 12} (1997) 002}, [\href{http://arxiv.org/abs/hep-th/9711053}{{\tt
  hep-th/9711053}}].

\bibitem{Lunin:2001jy}
O.~Lunin and S.~D. Mathur, \emph{{AdS / CFT duality and the black hole
  information paradox}},
  \href{http://dx.doi.org/10.1016/S0550-3213(01)00620-4}{\emph{Nucl. Phys. B}
  {\bf 623} (2002) 342--394}, [\href{http://arxiv.org/abs/hep-th/0109154}{{\tt
  hep-th/0109154}}].

\bibitem{Lunin:2002qf}
O.~Lunin and S.~D. Mathur, \emph{{Statistical interpretation of Bekenstein
  entropy for systems with a stretched horizon}},
  \href{http://dx.doi.org/10.1103/PhysRevLett.88.211303}{\emph{Phys. Rev.
  Lett.} {\bf 88} (2002) 211303},
  [\href{http://arxiv.org/abs/hep-th/0202072}{{\tt hep-th/0202072}}].

\bibitem{Mathur:2005zp}
S.~D. Mathur, \emph{{The Fuzzball proposal for black holes: An Elementary
  review}}, \href{http://dx.doi.org/10.1002/prop.200410203}{\emph{Fortsch.
  Phys.} {\bf 53} (2005) 793--827},
  [\href{http://arxiv.org/abs/hep-th/0502050}{{\tt hep-th/0502050}}].

\bibitem{Mathur:2008nj}
S.~D. Mathur, \emph{{Fuzzballs and the information paradox: A Summary and
  conjectures}},  \href{http://arxiv.org/abs/0810.4525}{{\tt 0810.4525}}.

\bibitem{Mathur:2009hf}
S.~D. Mathur, \emph{{The Information paradox: A Pedagogical introduction}},
  \href{http://dx.doi.org/10.1088/0264-9381/26/22/224001}{\emph{Class. Quant.
  Grav.} {\bf 26} (2009) 224001}, [\href{http://arxiv.org/abs/0909.1038}{{\tt
  0909.1038}}].

\bibitem{Bianchi:2017sds}
M.~Bianchi, D.~Consoli and J.~Morales, \emph{{Probing Fuzzballs with Particles,
  Waves and Strings}},
  \href{http://dx.doi.org/10.1007/JHEP06(2018)157}{\emph{JHEP} {\bf 06} (2018)
  157}, [\href{http://arxiv.org/abs/1711.10287}{{\tt 1711.10287}}].

\bibitem{Bianchi:2018kzy}
M.~Bianchi, D.~Consoli, A.~Grillo and J.~F. Morales, \emph{{The dark side of
  fuzzball geometries}},
  \href{http://dx.doi.org/10.1007/JHEP05(2019)126}{\emph{JHEP} {\bf 05} (2019)
  126}, [\href{http://arxiv.org/abs/1811.02397}{{\tt 1811.02397}}].

\bibitem{Bena:2018mpb}
I.~Bena, E.~J. Martinec, R.~Walker and N.~P. Warner, \emph{{Early Scrambling
  and Capped BTZ Geometries}},
  \href{http://dx.doi.org/10.1007/JHEP04(2019)126}{\emph{JHEP} {\bf 04} (2019)
  126}, [\href{http://arxiv.org/abs/1812.05110}{{\tt 1812.05110}}].

\bibitem{Bena:2019azk}
I.~Bena, P.~Heidmann, R.~Monten and N.~P. Warner, \emph{{Thermal Decay without
  Information Loss in Horizonless Microstate Geometries}},
  \href{http://dx.doi.org/10.21468/SciPostPhys.7.5.063}{\emph{SciPost Phys.}
  {\bf 7} (2019) 063}, [\href{http://arxiv.org/abs/1905.05194}{{\tt
  1905.05194}}].

\bibitem{Bianchi:2020des}
M.~Bianchi, A.~Grillo and J.~F. Morales, \emph{{Chaos at the rim of black hole
  and fuzzball shadows}},
  \href{http://dx.doi.org/10.1007/JHEP05(2020)078}{\emph{JHEP} {\bf 05} (2020)
  078}, [\href{http://arxiv.org/abs/2002.05574}{{\tt 2002.05574}}].

\bibitem{Bena:2015bea}
I.~Bena, S.~Giusto, R.~Russo, M.~Shigemori and N.~P. Warner, \emph{{Habemus
  Superstratum! A constructive proof of the existence of superstrata}},
  \href{http://dx.doi.org/10.1007/JHEP05(2015)110}{\emph{JHEP} {\bf 05} (2015)
  110}, [\href{http://arxiv.org/abs/1503.01463}{{\tt 1503.01463}}].

\bibitem{Bena:2016agb}
I.~Bena, E.~Martinec, D.~Turton and N.~P. Warner, \emph{{Momentum Fractionation
  on Superstrata}},
  \href{http://dx.doi.org/10.1007/JHEP05(2016)064}{\emph{JHEP} {\bf 05} (2016)
  064}, [\href{http://arxiv.org/abs/1601.05805}{{\tt 1601.05805}}].

\bibitem{Bena:2016ypk}
I.~Bena, S.~Giusto, E.~J. Martinec, R.~Russo, M.~Shigemori, D.~Turton et~al.,
  \emph{{Smooth horizonless geometries deep inside the black-hole regime}},
  \href{http://dx.doi.org/10.1103/PhysRevLett.117.201601}{\emph{Phys. Rev.
  Lett.} {\bf 117} (2016) 201601}, [\href{http://arxiv.org/abs/1607.03908}{{\tt
  1607.03908}}].

\bibitem{Bena:2017xbt}
I.~Bena, S.~Giusto, E.~J. Martinec, R.~Russo, M.~Shigemori, D.~Turton et~al.,
  \emph{{Asymptotically-flat supergravity solutions deep inside the black-hole
  regime}}, \href{http://dx.doi.org/10.1007/JHEP02(2018)014}{\emph{JHEP} {\bf
  02} (2018) 014}, [\href{http://arxiv.org/abs/1711.10474}{{\tt 1711.10474}}].

\bibitem{Bianchi:2017bxl}
M.~Bianchi, J.~F. Morales, L.~Pieri and N.~Zinnato, \emph{{More on microstate
  geometries of 4d black holes}},
  \href{http://dx.doi.org/10.1007/JHEP05(2017)147}{\emph{JHEP} {\bf 05} (2017)
  147}, [\href{http://arxiv.org/abs/1701.05520}{{\tt 1701.05520}}].

\bibitem{Bena:2017upb}
I.~Bena, D.~Turton, R.~Walker and N.~P. Warner, \emph{{Integrability and
  Black-Hole Microstate Geometries}},
  \href{http://dx.doi.org/10.1007/JHEP11(2017)021}{\emph{JHEP} {\bf 11} (2017)
  021}, [\href{http://arxiv.org/abs/1709.01107}{{\tt 1709.01107}}].

\bibitem{Giusto:2009qq}
S.~Giusto, J.~F. Morales and R.~Russo, \emph{{D1D5 microstate geometries from
  string amplitudes}},
  \href{http://dx.doi.org/10.1007/JHEP03(2010)130}{\emph{JHEP} {\bf 03} (2010)
  130}, [\href{http://arxiv.org/abs/0912.2270}{{\tt 0912.2270}}].

\bibitem{Giusto:2011fy}
S.~Giusto, R.~Russo and D.~Turton, \emph{{New D1-D5-P geometries from string
  amplitudes}}, \href{http://dx.doi.org/10.1007/JHEP11(2011)062}{\emph{JHEP}
  {\bf 11} (2011) 062}, [\href{http://arxiv.org/abs/1108.6331}{{\tt
  1108.6331}}].

\bibitem{Bianchi:2016bgx}
M.~Bianchi, J.~F. Morales and L.~Pieri, \emph{{Stringy origin of 4d black hole
  microstates}}, \href{http://dx.doi.org/10.1007/JHEP06(2016)003}{\emph{JHEP}
  {\bf 06} (2016) 003}, [\href{http://arxiv.org/abs/1603.05169}{{\tt
  1603.05169}}].

\bibitem{Bianchi:2020bxa}
M.~Bianchi, D.~Consoli, A.~Grillo, J.~F. Morales, P.~Pani and G.~Raposo,
  \emph{{Distinguishing fuzzballs from black holes through their multipolar
  structure}},  \href{http://arxiv.org/abs/2007.01743}{{\tt 2007.01743}}.

\bibitem{Blanchet:2006zz}
L.~Blanchet, \emph{{Gravitational radiation from post-Newtonian sources and
  inspiralling compact binaries}}, {\emph{Living Rev. Rel.} {\bf 9} (2006) 4}.

\bibitem{Sotiriou:2004ud}
T.~P. Sotiriou and T.~A. Apostolatos, \emph{{Corrected multipole moments of
  axisymmetric electrovacuum spacetimes}},
  \href{http://dx.doi.org/10.1088/0264-9381/21/24/003}{\emph{Class. Quant.
  Grav.} {\bf 21} (2004) 5727--5733},
  [\href{http://arxiv.org/abs/gr-qc/0407064}{{\tt gr-qc/0407064}}].

\bibitem{Hartle:1968si}
J.~B. Hartle and K.~S. Thorne, \emph{{Slowly Rotating Relativistic Stars. II.
  Models for Neutron Stars and Supermassive Stars}},
  \href{http://dx.doi.org/10.1086/149707}{\emph{Astrophys. J.} {\bf 153} (1968)
  807}.

\bibitem{Bena:2007kg}
I.~Bena and N.~P. Warner, \emph{{Black holes, black rings and their
  microstates}},
  \href{http://dx.doi.org/10.1007/978-3-540-79523-0\_1}{\emph{Lect. Notes
  Phys.} {\bf 755} (2008) 1--92},
  [\href{http://arxiv.org/abs/hep-th/0701216}{{\tt hep-th/0701216}}].

\bibitem{Gibbons:2013tqa}
G.~Gibbons and N.~Warner, \emph{{Global structure of five-dimensional
  fuzzballs}},
  \href{http://dx.doi.org/10.1088/0264-9381/31/2/025016}{\emph{Class. Quant.
  Grav.} {\bf 31} (2014) 025016}, [\href{http://arxiv.org/abs/1305.0957}{{\tt
  1305.0957}}].

\bibitem{Bates:2003vx}
B.~Bates and F.~Denef, \emph{{Exact solutions for supersymmetric stationary
  black hole composites}},
  \href{http://dx.doi.org/10.1007/JHEP11(2011)127}{\emph{JHEP} {\bf 11} (2011)
  127}, [\href{http://arxiv.org/abs/hep-th/0304094}{{\tt hep-th/0304094}}].

\bibitem{Denef_2011}
F.~Denef and G.~W. Moore, \emph{Split states, entropy enigmas, holes and
  halos}, \href{http://dx.doi.org/10.1007/jhep11(2011)129}{\emph{Journal of
  High Energy Physics} {\bf 2011} (Nov, 2011) }.

\bibitem{Raeymaekers_2008}
J.~Raeymaekers, W.~V. Herck, B.~Vercnocke and T.~Wyder, \emph{5d fuzzball
  geometries and 4d polar states},
  \href{http://dx.doi.org/10.1088/1126-6708/2008/10/039}{\emph{Journal of High
  Energy Physics} {\bf 2008} (Oct, 2008) 039–039}.

\bibitem{Boer_2008}
J.~d. Boer, F.~Denef, S.~El-Showk, I.~Messamah and D.~V.~d. Bleeken,
  \emph{Black hole bound states inads3×s2},
  \href{http://dx.doi.org/10.1088/1126-6708/2008/11/050}{\emph{Journal of High
  Energy Physics} {\bf 2008} (Nov, 2008) 050–050}.

\bibitem{Bena:2020see}
I.~Bena and D.~R. Mayerson, \emph{{A New Window into Black Holes}},
  \href{http://arxiv.org/abs/2006.10750}{{\tt 2006.10750}}.

\bibitem{Yagi:2016bkt}
K.~Yagi and N.~Yunes, \emph{{Approximate Universal Relations for Neutron Stars
  and Quark Stars}},
  \href{http://dx.doi.org/10.1016/j.physrep.2017.03.002}{\emph{Phys. Rept.}
  {\bf 681} (2017) 1--72}, [\href{http://arxiv.org/abs/1608.02582}{{\tt
  1608.02582}}].

\bibitem{Hartle:1967he}
J.~B. Hartle, \emph{{Slowly rotating relativistic stars. 1. Equations of
  structure}}, \href{http://dx.doi.org/10.1086/149400}{\emph{Astrophys. J.}
  {\bf 150} (1967) 1005--1029}.

\bibitem{Ryan:1996nk}
F.~D. Ryan, \emph{{Spinning boson stars with large selfinteraction}},
  \href{http://dx.doi.org/10.1103/PhysRevD.55.6081}{\emph{Phys. Rev.} {\bf D55}
  (1997) 6081--6091}.

\bibitem{Yagi:2015upa}
K.~Yagi and N.~Yunes, \emph{{Relating follicly-challenged compact stars to bald
  black holes: A link between two no-hair properties}},
  \href{http://dx.doi.org/10.1103/PhysRevD.91.103003}{\emph{Phys. Rev. D} {\bf
  91} (2015) 103003}, [\href{http://arxiv.org/abs/1502.04131}{{\tt
  1502.04131}}].

\bibitem{Bena:2020uup}
I.~Bena and D.~R. Mayerson, \emph{{Black Holes Lessons from Multipole Ratios}},
   \href{http://arxiv.org/abs/2007.09152}{{\tt 2007.09152}}.

\bibitem{Townsend:2002yf}
P.~K. Townsend, \emph{{Surprises with angular momentum}},
  \href{http://dx.doi.org/10.1007/s00023-003-0915-0}{\emph{Annales Henri
  Poincare} {\bf 4} (2003) S183--S195},
  [\href{http://arxiv.org/abs/hep-th/0211008}{{\tt hep-th/0211008}}].

\end{thebibliography}\endgroup
  
 \end{document}